\documentstyle[amssymb,prb,aps,twocolumn,epsf]{revtex}

\begin{document}
\twocolumn[\hsize\textwidth\columnwidth\hsize\csname@twocolumnfalse\endcsname

\draft

\title{Melting and transverse depinning of driven vortex lattices
in the periodic pinning of Josephson junction arrays}
\author{Ver\'{o}nica I. Marconi and Daniel Dom\'{\i}nguez}
\address{ Centro At\'{o}mico Bariloche, 8400 San Carlos 
de Bariloche, R\'{\i}o Negro, Argentina}

\maketitle

\begin{abstract}
We study the non-equilibrium dynamical regimes  of a moving vortex lattice in 
the periodic pinning of a  Josephson junction array (JJA) for {\it finite
temperatures} in the case of a fractional or submatching field.  We obtain a 
phase diagram for the current driven JJA  as a function of the driving current
$I$ and temperature $T$.  We find that when the vortex lattice is driven by a
current, the depinning transition at $T_p(I)$  and the melting transition at
$T_M(I)$  become separated even for a field for which they coincide in
equilibrium. We also distinguish between the depinning of the vortex lattice in
the direction of the current drive, and the {\it transverse depinning} in the
direction perpendicular to the drive. The transverse depinning corresponds to
the onset of transverse  resistance in a moving vortex lattice at a given
temperature $T_{tr}$.  For driving currents  above the critical current  we
find that   the moving vortex lattice has first a transverse depinning
transition at low $T$,  and later a melting transition at a higher temperature,
$T_{M}>T_{tr}$.

\end{abstract}

]

\section{INTRODUCTION}

The behavior of superconducting vortices in the presence of periodic
pinning shows very rich static and dynamic
phenomena.\cite{martinoli,wire,jjarev,jjpin,jjafrus,dot,hole,sfield,yan,%
martinoli0,PT,martinoli2,nelson,franz,hattel,nori0,carneiro0,nori,nori2,%
md99,carneiro99,reich,carneiro,prbnos}
The competition between the repulsive vortex-vortex interaction and
the attractive periodic pinning potential results in novel vortex
structures at low temperatures.
\cite{jjarev,sfield,yan,martinoli0,nori0,carneiro0}
The equilibrium phase transitions
of these vortex structures and their various dynamical regimes
when driven out of equilibrium are of great interest both
experimentally\cite{martinoli,wire,jjarev,jjpin,jjafrus,dot,hole,sfield,%
yan,martinoli2} 
and theoretically.\cite{jjarev,martinoli0,PT,martinoli2,nelson,franz,hattel,%
nori0,carneiro0,nori,nori2,md99,carneiro99,reich,carneiro,prbnos} 
Several techniques have
been developed to fabricate
in superconducting samples an artificial periodic  pinning  structure: 
thickness modulated superconducting films,\cite{martinoli}
superconducting wire networks,\cite{wire} 
Josephson junction arrays,\cite{jjarev,jjpin,jjafrus}
magnetic dot arrays,\cite{dot} 
submicron hole lattices,\cite{hole,sfield}
and pinning induced by Bitter decoration.\cite{yan}
The ground states of these systems can be either commensurate or
incommensurate vortex structures depending on the vortex density
(i.e the magnetic field). In the commensurate case,
a ``matching'' field is defined 
when the number of vortices $N_v$ is an integer multiple of the number
of pinning sites $N_p$: $N_v=nN_p$. 
A ``submatching'' or ``fractional'' field is defined
when  $N_v$ is a rational multiple of $N_p$: $N_v=fN_p$ with $f=p/q$.
One of the main properties of periodic pinning is that there are
enhanced critical currents and resistance minima both
for fractional and for matching magnetic fields,
for which the vortex lattice is strongly pinned. 

In the case of Josephson junction arrays\cite{jjarev} (JJA)
the discrete lattice structure  of Josephson junctions induces
an effective periodic pinning potential
(the so-called ``egg-carton'' potential) which at low temperatures
confines the vortices at the centers of the unit cells of the
network.\cite{jjpin} There are strong commensurability effects
for submatching fields $f=p/q$, for which the vortices arrange
in an ordered $q\times q$ superlattice that is commensurate with
the underlying array of junctions. The transition temperature
$T_c(f)$ and the critical current $I_c(f)$ have maxima for 
rational $f=p/q$, which have been observed
experimentally.\cite{jjarev,jjafrus}
Moreover, as we will discuss here, 
the model that describes the physics of the JJA
can be thought as a discrete lattice London model for thin
film superconductors with periodic arrays of holes. 
However, this comparison can be valid only for low submatching fields
since it can not describe the effects of interstitial vortices.

The equilibrium 
phase transitions at finite temperatures of two dimensional 
systems with periodic pinning  have been studied in the past.
\cite{PT,martinoli2,nelson,franz,hattel} 
It is possible to have a depinning phase transition of
the commensurate ground states at a temperature $ T_{p} $ and a
melting transition of the vortex lattice at a temperature $ T_{M} $.
\cite{PT,martinoli2,nelson,franz,hattel}
Franz and  Teitel\cite{franz}  have studied
this problem for the case of submatching fields.
For $0<T<T_p$ there is  a pinned phase in which the vortex lattice
(VL) is pinned commensurably to the periodic potential
and has long-range order. For $T_p<T<T_M$ there is a floating
VL which is depinned and has quasi-long-range order. 
For high submatching fields ($f\gtrsim 1/30$ )
both transitions coincide, \( T_{p}=T_{M} \), while for low submatching
fields ($f\lesssim 1/30$) 
both transitions are different with \( T_{p}<T_{M}
\).\cite{franz,hattel} 

The non-equilibrium dynamics of driven 
vortex lattices interacting either with
random or periodic pinning shows 
an interesting  variety of behavior.
\cite{nori,nori2,md99,carneiro99,reich,carneiro,%
prbnos,KV,gld,bmr,pardo,simu,dgb,dd99} 
Many recent studies have concentrated in the 
problem of the driven VL in the presence of {\it random} pinning.
\cite{KV,gld,bmr,pardo,simu,dgb,dd99}
When there is a large driving current
the effect of the pinning potential is reduced, and the nature of the
fastly moving vortex structure has been under active discussion.
The moving vortex phase has been proposed to be either a crystalline
structure, a moving glass,  a moving smectic
 or a moving transverse glass. \cite{KV,gld,bmr} These
moving phases have been studied both experimentally \cite{pardo} and
numerically. \cite{simu,dgb,dd99} 
Motivated by these results, the dynamical regimes of  the
moving VL in the presence of {\it periodic} pinning has also become
a subject of interest.\cite{nori,nori2,md99,carneiro99,reich,carneiro,%
prbnos}
At zero temperature,  the dynamical phases of vortices driven
by an external current with a periodic array of pinning sites
has been studied in very detail by F. Nori and
coworkers.\cite{nori,nori2}
A complex variety of regimes has been found, particularly
for $N_v>N_p$ where the motion of interstitial vortices
leads to several interesting dynamical phases.

In this paper we study the dynamical regimes 
of a moving VL in  the periodic pinning of a 
Josephson junction array (JJA) for {\it finite temperatures}
in the case of a {\it submatching} field ($f=1/25$). 
We obtain a  phase
diagram as a function of the driving current $I$ and temperature $T$,
which is shown in Fig.~1. We find that when the VL is driven
by a low current, the depinning and melting transitions can become
separated even for a field for which they coincide in equilibrium.
Moreover, we can distinguish between the depinning of the VL in the
direction of the current drive, and the \textit{transverse depinning}
in the direction perpendicular to the drive. This later case
corresponds to the vanishing of the transverse critical current in a
moving VL at a given temperature $T_{tr}$, or equivalently, to the
vanishing of the transverse superconducting coherence. 
We obtain three distinct regimes at low temperatures:
(i){\it Pinned vortex lattice:} for $0<T<T_p(I)$ there is an ordered
VL which has crystaline long-range order, superconducting coherence
(i.e., a finite helicity modulus) 
and zero resistance  both in the longitudinal and transverse
directions.
(ii){\it Transversely pinned vortex lattice:} for $T_p(I)<T<T_{tr}(I)$
there is a moving VL which has anisotropic Bragg peaks, quasi-long range
order, transverse superconducting coherence and 
zero transverse resistivity. There is a finite
transverse critical current. This regime also
has strong orientational pinning effects\cite{prbnos} 
in the [1,0] and [0,1] lattice directions.
(iii){\it Floating vortex lattice:} for $T_{tr}(I)<T<T_M(I)$ there
is a moving VL which is unpinned in both directions and it has
quasi-long range crystalline order with a strong anisotropy.
Some of the results we discuss here were reported by us previously in
a shorter and less detailed version.\cite{md99}
After our work in Ref.~\onlinecite{md99}, 
further studies of thermal effects in a moving vortex
lattice with a periodic pinning array have  been
reported for $N_v=N_p$ \cite{reich}  and 
for $N_v>N_p$. \cite{carneiro99,carneiro}
Some of these results are similar to ours.

The remainder of our paper is organized as follows.
In Sec.~II we introduce the theoretical model used for the dynamics
of the JJA. We also discuss how  this model can be mapped to a superconducting
film with a periodic array of holes. 
In Sec.~III we discuss the melting transition near
equilibrium when the JJA is driven by a very small current.
In Sec.~IV we present our results for
the transport properties, analyzing the temperature dependence of the 
current-voltage curves and the onset of resistivity at $T_p$ for
different currents.
In Sec.~V we present our results for the transverse depinning
transitions at different temperatures and driving currents. We study
the transverse current-voltage characteristics both at $T=0$ and
at finite $T$. We define the transverse depinning temperature $T_{tr}$
from the behavior of the transverse resistivity.
In Sec.~VI we present in detail our results for the different dynamical
regimes. For various fixed driving currents, we analyze the behavior
of structure factor, longitudinal and transverse
resistance and helicity modulus through the different regimes
as a function of temperature.
These results constitute the central core of this paper, from which the 
phase diagram of Fig.~1 was obtained.
In Sec.~VII we discuss the orientational pinning effects, which
are observed when the direction of the driving current is 
rotated with respect to the JJA square lattice.
Finally in Sec.~VIII we present our discussion and conclusions.
We also add an  Appendix  at the end, where we provide  a detailed definition
of the adequate periodic boundary conditions for a JJA with an external
magnetic field and an external driving current, as well as the
algorithm used for the numerical simulation.

\section{MODEL AND DYNAMICS}

\subsection{RSJ dynamics}

We consider a square Josephson junction array with $L\times L$
superconducting nodes. The   nodes are in the lattice 
sites ${\bf n}=(n_{x},n_{y})$ and their 
superconducting phases are $\theta ({\bf n})$.
We study the dynamics of JJA using the resistively shunted 
junction (RSJ)
model for the junctions of the square
network.\cite{dyna0,falo,eik,acvs,kim93%
vorstroud,vorjose}
In this case, the current
flowing in the junction between two superconducting nodes in the JJA is
modeled as the sum of the Josephson supercurrent and the normal current: 
\begin{equation}
I_{\mu }({\bf n})=I_{0}\sin \theta _{\mu }({\bf n})+\frac{\hbar}{2e
R_{N}}\frac{d\theta _{\mu }({\bf n})}{dt}+\eta _{\mu }({\bf %
n},t)
\label{rsj}
\end{equation}
where $I_{0}$ is the critical current of the junction between the sites $%
{\bf n}$ and ${\bf n}+{\bf \mu }$,  (${\bf \mu }=
{\bf \hat{x}},{\bf \hat{y}}$), $R_{N}$ is the normal state
resistance and 
\begin{equation}
\theta _{\mu }({\bf n})=\theta ({\bf n}+{\bf \mu })-\theta ({\bf n})-A_{\mu
}({\bf n})=\Delta _{\mu }\theta ({\bf n})-A_{\mu }({\bf n})
\end{equation}
is the gauge invariant phase difference with 
\begin{equation}
A_{\mu }({\bf n})=\frac{2\pi }{\Phi _{0}}\int_{{\bf n}a}^{({\bf n}+{\bf \mu }%
)a}{\bf A}\cdot d{\bf l}.
\end{equation}
The thermal noise fluctuations $\eta _{\mu }$ have correlations 
\begin{equation}
\langle \eta _{\mu }({\bf n},t)\eta _{\mu ^{\prime }}({\bf n^{\prime }}%
,t^{\prime })\rangle =\frac{2kT}{R_{N}}\delta _{\mu ,\mu ^{\prime }}\delta _{%
{\bf n},{\bf n^{\prime }}}\delta (t-t^{\prime })
\end{equation}

In the presence of an external magnetic field $H$ we have 
\begin{eqnarray}
\Delta_{\mu}\times A_{\mu}({\bf n})&=&A_x({\bf n})-A_x({\bf n}+{\bf y})+ 
A_y({\bf n}+{\bf x})-A_y({\bf n})\nonumber\\
&=&2\pi f,
\end{eqnarray}
$f=H a^2/\Phi_0$ and $a$ is the array lattice spacing. We take periodic
boundary conditions (p.b.c) in both directions in the presence of an
external current ${\bf I}=I{\bf \hat{y}}$ in arrays with $L\times L$ 
junctions.\cite{dd99} (See the Appendix).
The vector potential is taken as 
\begin{equation}
A_{\mu}({\bf n},t)=A_{\mu}^0({\bf n})-\alpha_{\mu}(t)
\end{equation}
where in the Landau gauge $A^0_x({\bf n})=-2\pi f n_y$, $A^0_y({\bf n})=0$
and $\alpha_{\mu}(t)$ allows for total voltage fluctuations under periodic
boundary conditions. 
In this gauge the p.b.c. for the phases are:\cite{md99,dd99} 
\begin{eqnarray}
\theta(n_x+L,n_y)&=&\theta(n_x,n_y)  \nonumber \\
\theta(n_x,n_y+L)&=&\theta(n_x,n_y)-2\pi f Ln_x.
\end{eqnarray}
 We also consider local conservation of current, 
\begin{equation}
\Delta_\mu\cdot I_{\mu}({\bf n})=\sum_{\mu} I_{\mu}({\bf n})- I_{\mu}({\bf n}%
-{\bf \mu})=0.
\label{divI}
\end{equation}
After Eqs.~(\ref{rsj},\ref{divI}) 
we obtain the following  equations for the phases,\cite{md99,dd99} 
\begin{equation}
\frac{\hbar}{2eR_{N}}\Delta_{\mu}^2\frac{d\theta({\bf n})}{dt}=
-\Delta_{\mu}\cdot [S_{\mu}({\bf n})+\eta_{\mu}({\bf n},t)]
\label{dyn0}
\end{equation}
where 
\begin{equation}
S_{\mu}({\bf n})=I_0\sin[\Delta_\mu\theta({\bf n})-A_{\mu}^0({\bf n})-
\alpha_{\mu}]\;,
\end{equation}
and the discrete Laplacian is 
\begin{eqnarray}
\Delta^2_\mu f({\bf n})&=&f({\bf n}+{\bf \hat x}) +f({\bf n}-%
{\bf \hat x})+f({\bf n}+{\bf \hat y}) +f({\bf n}-{\bf \hat y}%
)\nonumber\\
& &\;-4\,f({\bf n}).
\end{eqnarray}
The Laplacian can be inverted with the square lattice Green's function
$G_{{\bf n},{\bf n'}}$:
\begin{equation}
 \Delta^2_\mu G_{{\bf n},{\bf n'}}=\delta_{{\bf n},{\bf n'}}.
\label{green}
\end{equation}

Since we take periodic boundary conditions (see Appendix),
the  total current has to be fixed by: 
\begin{eqnarray}
I_x&=&\frac{1}{L^2}\left[\sum_{%
{\bf n}} I_0\sin\theta_x({\bf n})+\eta_x({\bf n},t)\right] + \frac{\hbar}{%
2eR_N} \frac{d\alpha_x}{dt}\;,  \nonumber \\
& & \label{tot}\\
I_y&=&\frac{1}{L^2}\left[\sum_{%
{\bf n}} I_0\sin\theta_y({\bf n})+\eta_y({\bf n},t)\right] + \frac{\hbar}{%
2eR_N} \frac{d\alpha_y}{dt}\;,\nonumber 
\end{eqnarray}
These equations determine the dynamics of $\alpha_\mu(t)$. \cite{dd99} 
For this case we take $I_x=0$ and $I_y=I$.
After Eqs.~(\ref{dyn0},\ref{green},\ref{tot}) we obtain 
the following set of dynamical equations,\cite{md99,dd99} 
\begin{eqnarray}
\frac{d\theta({\bf n})}{dt}&=&-\sum_{\bf n'}G_{{\bf n},{\bf n'}}\Delta_{\mu}%
\cdot \left[S_{\mu}({\bf n'})+\eta_{\mu}({\bf n'},t)\right],\label{dyn1} \\
\frac{d\alpha_{\mu}}{dt}&=&I_{\mu} -\frac{1}{L^2}\sum_{{\bf n}%
} S_{\mu}({\bf n})+\eta_{\mu}({\bf n},t),\label{dyn2}
\end{eqnarray}
where we have normalized currents by $I_0$, 
time by $\tau_J=2eR_{N}I_0/\hbar$, 
and temperature by $I_0\Phi_0/2\pi k_B$.

\subsection{Comparison with thin film with a periodic array of holes}

Let us consider a superconducting thin film with a square array of
holes, which act as pinning sites for vortices. There are $N_p=L^2$
pinning sites separated by a distance $a$. 
The current density in the superconducting film is given by
 the sum of the supercurrent and the normal current:
\begin{eqnarray}
{\bf J}&=&{\bf J}_S+{\bf J}_N\nonumber\\
{\bf J}&=&\frac{ie\hbar}{m^*}
\left[\Psi^*{\bf D}\Psi-({\bf D}\Psi)^*\Psi\right]+
\frac{\sigma\Phi_0}{2\pi c}\frac{\partial}{\partial
t}\left(\nabla\theta-\frac{2\pi}{\Phi_0}{\bf A}\right)
\end{eqnarray}
with ${\bf D}=\nabla+i\frac{2\pi}{\Phi_0}{\bf A}$, 
$\Psi({\bf r})=|\Psi({\bf r})|\exp[i\theta({\bf r})]$  the
superconducting order parameter and $\sigma$  the normal state
conductivity. These equations are valid everywhere
in the film except in the hole regions. If the number of vortices
$N_v=BL^2a^2/\Phi_0$ is much smaller than the number of pinning sites
$N_p$, all vortices will be centered in the holes in equilibrium. 
In this case we can assume that $|\Psi({\bf
n})|\approx|\Psi_0|$  is homogeneous in the superconducting film.
Therefore the dynamics is given by the superconducting phase 
$\theta({\bf r})$,  corresponding to a London model in a sample with holes.
After considering current conservation
$ \nabla\cdot{\bf J}=0,$ we obtain the London dynamical equations 
for the phases
in this multiply connected geometry. Since $N_v\ll N_p$, we make the 
approximation of solving the equations in a discrete grid of spacing $a$.
This means that we take as the relevant dynamical variables the phases 
$\theta({\bf r}_{\bf n})$ defined
in the sites which are dual to the pinning sites. They represent
the average superconducting phase in each
superconducting square defined by four pinning sites. 
Therefore, we  take the discretization ${\bf r}_{\bf
n}=(n_xa,n_ya)=a{\bf n}$. [Pinning sites at centered
at positions ${\bf r}_{\bf p}=(n_x + 1/2 , n_y+1/2)a$]. 
The derivatives in the supercurrent are discretized in a gauge-invariant way as
\begin{equation}
D_\mu \Psi({\bf r})\rightarrow\frac{1}{a}\left[\Psi({\bf n}+
{\bf\mu})-e^{-i2\pi A_\mu({\bf n})/\Phi_0}\Psi({\bf n})\right].
\end{equation}
After doing this, we obtain an equation analogous to (\ref{rsj}).
 Now $I_\mu({\bf n})$ has to be interpreted as current density
normalized by $J_0=2e\hbar |\Psi_0|^2/ma=\Phi_0/(8\pi^2\lambda^2a)$,
time normalized by $\tau=c/(4\pi\sigma\lambda^2)$,
and the fraction of vortices is  $f=N_v/N_p=Ba^2/\Phi_0$.
This leads to a set of dynamical equations of the same form as 
Eqs.(\ref{dyn1},\ref{dyn2}). Therefore, we expect that for $f\ll 1$ the
model for a JJA also gives a good representation of the physics
of a superconducting film with a square array of holes
(meaning that effects of interstitial vortices are neglected
for $N_v\ll N_p$). In other words, we expect that for a low
density of vortices the specific shape of the periodic pinning
potential (being either an egg-carton  or an  array of holes)
will not be physically relevant.

\subsection{Quantities calculated and simulation parameters}

The Langevin dynamical equations (\ref{dyn1},\ref{dyn2}) 
are solved with a second order
Runge-Kutta-Helfand-Greenside algorithm with time step $\Delta
t=0.1\tau_J$. The discrete periodic Laplacian is inverted with 
a fast Fourier + tridiagonalization algorithm as in Ref.\onlinecite{acvs}.
(See also the Appendix).

We calculate the following physical quantities:

(i) {\it Transverse superconducting coherence}: We obtain the helicity modulus
$\Upsilon_x$ in the direction transverse to the current as
\begin{eqnarray}
\Upsilon_x&=&\frac{1}{L^2}\left\langle\sum_{{\bf n}}\cos\theta_x({\bf
n})\right\rangle-\frac{1}{TL^4}\left\{\left\langle \left[\sum_{{\bf n}}
\sin\theta_x({\bf n})\right]^2\right\rangle\right. \nonumber\\ 
& & \left.- \left\langle \left[\sum_{{\bf n}}\sin\theta_x({\bf
 n})\right]\right\rangle^2\right\}\nonumber
\end{eqnarray}
Whenever we calculate the helicity modulus along $x$, we enforce
strict periodicity in $\theta$ by fixing $\alpha_x(t)=0$. 
(See Appendix).
 
(ii){\it Transport}: We calculate the transport response of the JJA
 from the time average of the total voltage as 
\begin{eqnarray}
V_x&=&\langle v_x(t)\rangle= \langle d\alpha_x(t)/dt\rangle  \nonumber \\
V_y&=&\langle v_y(t)\rangle= \langle d\alpha_y(t)/dt\rangle
\end{eqnarray}
with voltages normalized by $R_ {N}I_0$.

(iii) {\it Vortex structure}: We obtain the vorticity
at the plaquette ${\bf \tilde n}=(n_x+1/2,n_y+1/2)$ 
(associated to the site ${\bf n}$) as\cite{dj96} 
\begin{equation}
b({\bf \tilde n})=-\Delta_\mu\times{\rm nint}[\theta_\mu({\bf n})/2\pi]
\end{equation}
with ${\rm nint}[x]$ the nearest integer of $x$. We calculate the average
vortex structure factor as
\begin{equation}
S({\bf k})=\left\langle\left|\frac{1}{L^2}\sum_{\bf \tilde n}
b({\bf \tilde n})\exp(i{\bf k}\cdot{\bf \tilde n})\right|^2\right\rangle.
\end{equation}

\section{Transition near equilibrium}

We study JJA with a low magnetic field  corresponding to $f=1/25$, for
different system sizes of $L\times L$ junctions, with $L=50,100,150$.
Most of the results are for $L=100$, except when it is explicitly specified, 
and for $N_t=10^5$ iterations after a transient of $N_t/2$ iterations.

The ground state vortex configuration for $f=1/25$  is a tilted 
square-like vortex lattice (VL),\cite{teitel25} see Fig.2(a). We find
that this state
is stable for low currents and low temperatures (in fact, the structure
of Fig.2(a) corresponds to $I=0.01$ and $T=0.01$). The lattice is
oriented in the $[4a,3a]$ direction and commensurated with the underlying
periodic pinning potential of the square JJA.
The structure factor $S({\bf k})$ has the corresponding Bragg
peaks at wavevectors ${\bf G}$ in the reciprocal space, 
as can be seen in Fig.1(b). 
When the temperature is increased, the VL tends to disorder and above
the melting temperature
$T_M$ a random vortex array with
a liquid-like structure factor is obtained, Figs. 2(c) and
2(d).

We find a single equilibrium phase transition ($I=0$)
at $T_M\approx0.050\pm0.003$, which is in agreement with
the melting temperature obtained by Franz and Teitel\cite{franz} 
and Hattel and Wheatley\cite{hattel} for $f\gtrsim 1/30$.

We now apply a  very low current,  $I=0.01$, in order to
study the near-equilibrium transport response simultaneously with other
quantities like structure factor and helicity modulus.   We find a
phase transition at a temperature $T_M(I)\approx0.046\pm0.001$, which
is slightly lower than the equilibrium transition. In Fig.3(a) we see
that there is a large jump in the resistance $R=V/I$  at $T_c$, in good
agreement with the first-order nature of the equilibrium transition.
\cite{franz} 
The onset of resistivity is a signature of a depinning
transition in the direction of the drive.  This occurs simultaneously
with a melting of the vortex lattice, corresponding to the vanishing of
Bragg peaks,  as shown in Fig.3(b) for the two first reciprocal lattice
vectors  ${\bf G_1}=\frac{2\pi}{a}(-4/25,-3/25)$ and ${\bf
G_2}=\frac{2\pi}{a}(-3/25,4/25)$.
The size dependence of $S({\bf G}, T)$ is shown in the inset of
Fig.3(b).  In the direction of the current drive the helicity modulus
is ill-defined since total phase fluctuations are allowed. 
(See the Appendix).  However, in
the perpendicular direction to the drive the helicity modulus
$\Upsilon_x$ can be calculated, and it is a measure of the transverse
superconducting coherence. As we can see in Fig.3(c), transverse
superconductivity also vanishes at $T_M(I)$. Above $T_M$, we find that
the $\Upsilon_x(T)$ has large fluctuations around zero.

\section{Transport properties}

Let us now study the transport properties  
for larger currents. We calculate the current-voltage (IV)
characteristics for  different temperatures as well as the dc resistance
$R=V/I$ as a function of temperature for finite currents.

The zero temperature IV curve has a critical current, $I_c(0)=
0.114\pm0.002$, see Fig.4, which corresponds to the single vortex
depinning current in square JJA, \cite{jjpin} with the typical square
root depence at the onset. Similar behavior has been reported for zero
temperature IV curves for low values of $f$.\cite{vorstroud,vorjose}  
Above $I_c(0)$ there is an
almost linear  increase of voltage, corresponding to a ``flux flow''
regime, where there is a fastly moving VL. The structure factor of the
$T=0$ moving VL is the same as the corresponding one of the pinned VL
[Fig.2(a)]. The presence of periodic boundary conditions in our case 
prevents the occurrence of random or chaotic vortex motion  near the
critical current, as reported in early simulations with free boundary
conditions.\cite{dyna0,falo} 
In what follows we will restrict  our analysis for currents
$I<0.4$, where the collective behavior of the VL is  the dominant
physics. (At $I\sim 1$ there is a sharp increase of voltage when all
the junctions become normal, and $V\sim R_NI$ for $I\gg 1$).

The IV curves for finite temperatures are shown in Fig.5. For
temperatures below $T_M$ there is a nonlinear sharp rise in voltage 
which defines the apparent critical current $I_c(T)$. For example, we
can obtain this $I_c(T)$ with a  voltage criterion, which we choose as
$V < \frac{1}{N_t\Delta t}=10^{-4}$. In this case, we find that
$I_c(T)$ decreases with $T$, vanishing at $T_M$. It is interesting to
point out that  all the IV curves for different temperatures have a
crossing point  at $I^*=0.165$, see Fig.5(a). A crossing in the IVs 
has also  been reported in experiments in amorphous thin
films.\cite{hellerq} For
temperatures $T>T_M$ the IV curves tend to linear resistivity for low
currents. This is shown in Fig.5(b) in a log-log plot of $R(I)=V/I$  vs
$I$, where we see that $R(I)$  tends to a low current finite value for
$T>T_M$, while it has a strong nonlinear decrease for $T<T_M$.

Let us now study the dc resistance $R=V/I$ as a function of temperature
for a given applied current in the $y$-direction (Fig.6).  We start
with the perfectly ordered VL as an initial condition at $T=0$ and then
we slowly increase the temperature, keeping $I$ constant.  For currents
below the $T=0$ critical current, $I<I_c(0)$, the dc resistance is
negligibly small at low $T$, and it has a steep increase at a depinning
temperature $T_p(I)$, corresponding to the onset of vortex  motion. The
depinning temperature decreases for increasing currents, and the values
of $T_p(I)$ are coincident with the apparent critical currents $I_c(T)$
obtained from the IV curves.  For currents higher than $I_c(0)$, there
is always a large and  finite voltage for any temperature.  If 
$I_c(0)<I<I^*$, the $R(I)$ increases slightly with $T$ tending to a
constant value for  large $T$, while for $I>I^*$ the $R(I)$ decreases
with $T$.

\section{Transverse depinning}

What is its response to a
small current in the transverse direction when the driven vortex lattice is 
moving? Is the vortex lattice still
pinned in the transverse direction? Is there a transverse critical 
current for a moving VL?    The idea of a transverse depinning current
was introduced by Giamarchi and Le Doussal in Ref.\onlinecite{gld} for
moving vortex systems with random pinning at  zero temperature. The
possibility of such a critical current was later questioned by Balents,
Marchetti and Radzihovsky,\cite{bmr} where it was shown that this is
not true for any  finite temperature in {\it random pinning}; however a
strong nonlinear increase of the transverse voltage was predicted at an
``effective'' transverse critical current. In the case of {\it periodic
pinning} it is  more clear that a transverse critical current will
exist at $T=0$ since it is a commensurability effect. This has been
found in the $T=0$ simulation work of Reichhardt {\it et
al.}.\cite{nori}  It is also possible that this transverse critical
current will still be non-zero at $T\not=0$ in {\it periodic pinning}.
In fact, we have found in our previous work \cite{md99} that there is a
{\it thermal} transverse depinning in a {\it periodic} system, and we
will now analyze this behavior in detail.

First, a high longitudinal current $I>I_c(0)$ is applied at zero
temperature. Then, a current $I_{tr}$ is applied in the transverse
direction. In this way, a transverse current-voltage characteristics
can be obtained for each $I$.  This is shown in Fig.7(a) for
$I=0.16$. We clearly see that there is a finite transverse critical
current $I_{c,tr}\approx 0.12$, which is of the order of the single
vortex pinning barrier. We also show in Fig.7(a) how the longitudinal
voltage $V$ changes when the transverse current is varied. 
For $I_{tr}<I_{c,tr}$ the longitudinal voltage $V$ is almost constant.
At $I_{c,tr}$ there is a fast decay of $V$. When $I_{tr}=I$ we also
have $V=V_{tr}$ as expected for a drive at a degree of $\pi/4$.
Later, for $I_{tr}>I$ the vortex lattice becomes pinned in the other
direction, since now the  directions of ``longitudinal'' and
``transverse'' current are interchanged.

 Another possible measurement is to study
thermal transverse depinning. In this case, we start with a
longitudinal current  $I$, then a small transverse current is
applied,  $I_{tr}\ll I_{c,tr}$, and the temperature is slowly
increased. In this way, we can measure a transverse resistance
$R_{tr}=V_{tr}/I_{tr}$. In Fig.7(b) we plot this result for $I=0.16$
and $I_{tr}=0.01$. We find that for finite low temperatures $R_{tr}$ is
negligibly small within our numerical accuracy and it has a clear onset
at a  transverse critical temperature $T_{tr}$.

Let us now see how these results depend on the longitudinal current
$I$ and temperature $T$. We have calculated the transverse IV (Tr-IV)
curves  for different $I$ and $T$.  In Fig.8(a) we show the Tr-IV
curves for $I=0.06$ (low current regime) and in Fig.8(b) for
$I=0.16$ (high current regime). In  both cases, there is a clear
change of behavior in the Tr-IV curves when going through  a
characteristic transverse critical temperature $T_{tr}(I)$. For low
temperatures  there is a transverse critical  current  which tends to
vanish when $T$ aproachs $T_{tr}(I)$ from below.  In contrast, for
$T>T_{tr}$ there is a linear resistivity behaviour.

The transverse resistivity $R_{tr}=V_{tr}/I_{tr}$ as a function of
temperature  was calculated for different longitudinal currents
(Fig.9).  In all cases there is an onset of transverse response at a
given  temperature $T_{tr}(I)$. At low currents, $I<I_c(0)$, the
transverse depinning temperatures are almost constant,
$T_{tr}\sim0.02$  tending to increase slowly with $I$, see Fig.9(a). 
On the other hand, for  $I > I_c(0)$ the transverse depinning
temperatures  increase clearly with $I$, see Fig.9(b).

\section{Non-equilibrium regimes}

We will now study the different non-equilibrium regimes of vortex
driven lattices and characterize their possible dynamical transitions. 
The approach we will follow in this Section 
is to have a fixed current applied in the system 
and vary the temperature.
In this way, we look for the possible transitions
as a function of $T$ in a similar way as was done near equilibrium in
Sec.III.  

A few similar studies were done previously in related systems.
In Ref.~\onlinecite{dgb} the melting transition of a moving
vortex lattice in the three dimensional XY model was studied
in this way. 
In this work a first order transition was found as a function of
temperature in a strongly driven vortex lattice. 
In Ref.~\onlinecite{kim93} a current driven
two dimensional JJA at zero field was studied. 
The possibility of a transition  as a function of temperature
for finite  currents below the critical current was analyzed
in this case. 

In the following, we will separate our study in  three ranges of
current: (i) low currents, $I < 0.04$, 
(ii) intermediate currents $0.04< I < I_c(0)$, and high currents
$I > I_c(0)$.

\subsection{Low currents}

We show the results for a low current in Fig.10 for $I=0.03$. The
longitudinal dc response, $V/I$, is negligibly small at low $T$
and later it has a sharp increase of two orders of magnitude: this
defines the {\it depinning temperature}, $T_p$ [Fig.10(a)]. At higher
temperatures, $T>T_p$, the resistance is weakly $T$-dependent. 

Below $T_p$ the vortex lattice is pinned and its structure is similar
to the ground state: a vortex lattice commensurate with the underlying
square array and tilted in the $[4a,3a]$ direction.  Above $T_p$  the
vortex lattice is moving and it has an anisotropic structural order.   
If we analyze the structure factor in two different reciprocal lattice
directions $S(G_1)$ and $S(G_2)$, 
we can see this clearly (Fig.10(b)). 
Below $T_p$ the VL structure is isotropic and
$S(G_1)=S(G_2)$. Right above $T_p$ the height of the peaks decreases
with temperature, and the structure of the depinned VL is clearly
anisotropic,  $S(G_1)\not=S(G_2)$. Finally, at a melting temperature
$T_M$ the peaks vanish, and the vortex lattice melts into a vortex
liquid. In  Fig.11 we compare the behavior of the Bragg peaks
for two system sizes $L=50,100$. 
We see that the $S(G_{1,2})$ are size independent below $T_p$ as it should be
expected for a pinned phase.\cite{franz} 
For large temperatures $T>T_M$, in the liquid phase, the value
of $S(G_{1,2})$ is strongly size dependent, 
since it should go as $L^{-2}$.\cite{franz}
On the other hand,
for $T_p<T<T_M$ we find that the intensity of the Bragg
peaks is weakly dependent on system size.  
The clear change of behavior of the size dependence gives a 
good criterion to determine  $T_M$ (Fig.~11).

The helicity modulus in the
direction perpendicular to the current, $\Upsilon_x$,
decreases very slowly for $T<T_p$. Above $T_p$,
$\Upsilon_x$ has a faster decay with important
fluctuations  and tends to vanish at  $T_M$. 
For $T>T_M$, $\Upsilon_x$ oscillates around zero.

Therefore, for a small finite current the depinning and melting
transitions become separated with $T_p < T_M$.

\subsection{Intermediate currents}

At intermediate currents, $0.04<I<I_c(0)$, a {\it new} transition
appears: the {\it transverse depinning} of the  moving vortex lattice.
As discussed in Sec.V, one can measure transverse depinning by applying
a small transverse current while the VL is driven with a fixed 
longitudinal current.
This is shown in Fig. 12(a) for the case 
of $I=0.06$ and a small transverse
current, $I_{tr}=0.01$. We see that
there is an onset of transverse voltage  at $T_{tr}=0.02$.
We can also see that this transverse depinning temperature is
above the depinning temperature $T_p$ for longitudinal
resistance [Fig.12(b), $T_p=0.013$], and below 
the melting temperature $T_M$ for the vanishing of the Bragg
peaks [Fig.12(c), $T_M=0.037$]. 
Therefore, this transition occurs at an intermediate
temperature between the depinning
and the melting transitions, $T_p<T_{tr}<T_M$.
We also show in Fig.12(d) that 
the helicity modulus  begins to fall down slowly 
at $T_{tr}$, while for $T_{tr}<T<T_M$ it has strong fluctuations,
being difficult to interpret its behavior in this case.

It is interesting to study in more detail the behavior of 
the structure factor through all these transitions.
The temperature dependence of the intensity of the Bragg peaks
is shown in Fig.12(c). 
In Fig.13 we show examples of the structure factor $S({\bf k})$ 
at temperatures in the different regimes.
In the pinned phase, the $S({\bf k})$ is nearly the same as in
the ground state with delta-like Bragg peaks, see Fig.13(a).
For the moving VL, we can see that 
there is less anisotropy  in the {\it transversely pinned}
regime $T_p<T<T_{tr}$  [Fig.13(b)]
than in the {\it floating} regime $T_{tr}<T<T_M$ [Fig.13(c-d)].
Also in the inset of Fig.12(c) 
one can see that the intensity of the Bragg
peaks has a greater dependence with size for 
$T_{tr}<T<T_M$ when compared with the $T_p<T<T_{tr}$ regime.
Moreover, near $T_M$ the VL structure becomes  strongly anisotropic,
with the peak at ${\bf G}_1$ much larger than the peak at ${\bf G}_2$,
see Fig.13(d).

The anisotropy of the Bragg peaks of the moving VL  (in the regimes at
$T_p<T<T_M$) has two  characteristics: (i) the {\it width} of the peaks
increases  with $T$ in the direction of the applied current (the direction
perpendicular to the vortex motion), and (ii)  the {\it height} of the peaks
decreases in the direction of vortex motion. This can be observed in the
sequence of structure factors shown in Fig.13(b-d). The details of the
broadening of the Bragg  peaks  are shown in Fig.14(a) for the transversely
pinned regime  ($T_p<T=0.018<T_{tr}$) and in Fig.14(b) for the floating solid
regime ($T_{tr}<T=0.023<T_M$).

In Fig.15  we show the temperature dependence of
$S({\bf G_1})$ and $S({\bf G_2})$ for different currents.
In all the cases the three different regimes in temperatures
are observed: (i) pinned regime for $0<T<T_{p}$ with isotropic
and size independent Bragg peaks, (ii) transversely pinned
regime for $T_p < T < T_{tr}$ with weakly anisotropic Bragg peaks,
and (iii) the floating solid regime for  $T_{tr} < T < T_M$ with
strong anisotropy and size dependence in the Bragg peaks.

\subsection{High currents}

In the case of  currents  larger than $I_c(0)$, the vortex lattice
is already depinned at $T=0$. As we have seen in Sec.~V, this
zero-temperature moving VL has a finite transverse critical current,
and therefore it is pinned in the transverse direction.
When we slowly increase temperature from this state, we find
that the transverse resistive response is negligible for 
finite low temperatures. At a  temperature $T_{tr}$ there is a
jump to a finite transverse resistance $R_{tr}=V_{tr}/I_{tr}$.
For example,  this is shown for  $I=0.16$ with a small transverse current,
$I_x=0.01$ in Fig.16(a). 
The vortex lattice has an anisotropic structural order for all temperatures,
{\it i.e}, $S({\bf G_1})\neq S({\bf G_2})$,
and the height of the Bragg peaks vanishes at $T_M$, Fig.16(b). 
In the inset we show $S({\bf G_2})$ for different sizes, $L=50,100$, 
and we see that $T_M$ is size independent. 
Similar behaviour is found for $S({\bf G_1})$. 
The transverse helicity modulus $({\Upsilon_x})$  is 
almost constant for low
temperatures and starts to decrease at $T_{tr}$ presenting strong fluctuations 
for $T>T_{tr}$ , Fig16(c).

The analysis of the heights of the peaks in $S({\bf k})$ as a function of
system size $L$ is a good indicator of the translational correlations in the
system. This dependence is well known for two dimensional lattices. 
\cite{franz} For a pinned solid $S({\bf G})\sim1$, 
for a floating solid, $S({\bf G})\sim
L^{-\eta_G(I,T)}$ with $0<{\eta_G(I,T)}<2$, being this dependence a signature
of quasi-long range order, 
and for a normal liquid $S({\bf G})\sim L^{-2}$.  To
assure the existence of algebraic translational correlations, 
we have done 
this scaling study for currents $0.02<I<0.2$ and different temperatures
for system sizes of $L=50,100,150$. 
For the cases corresponding to the pinned regime ($T<T_p$)
we found $\eta_G\approx0$, as expected. 
In the transversely pinned regime, we show a case in Fig.17(a),
we find a power law fitting with  very small values of ${\eta_G}(I,T)$.
In the floating regime, we find larger values of $\eta_G$,
we show a case in Fig. 17(b).
In all the cases we have obtained that 
 $0<{\eta_G}(I,T)<2$ for $T_p<T<T_M$. 
Therefore, this finite size analysis shows the existence of 
quasi-long range order in the moving VL.
Also, we find that ${\eta_{G_1}} > {\eta_{G_2}}$ for all currents
and temperatures.   
The power-law exponent ${\eta_G}$ can be studied
at a constant current,  as a function of temperature. 
This is shown in Fig.18, for $I=0.16$
and different reciprocal lattice vectors. The exponent ${\eta}$ is
finite for the complete temperature range.
For $T<T_{tr}$ it has a small value $\eta_G\approx 0.01$. 
It has a fast increase near $T_{tr}$, 
where there is also a clear difference between ${\eta_{G_1}}$
and ${\eta_{G_2}}$.
Finally, it reaches a  value of  ${\eta=2}$ near $T=T_{M}$.

\section{Orientational pinning effects}

A very interesting  characterization of the different regimes can
be obtained by  studying  the effects of varying current direction.
\cite{prbnos,stroud99,choi} 
We apply a current $I$ at an angle $\phi$ with respect to the $[10]$
lattice direction, 
\begin{eqnarray}
I_x&=&I\cos\phi  \nonumber \\
I_y&=&I\sin\phi.
\end{eqnarray}
We study the voltage response when varying the orientation $\phi$ of the drive
while keeping fixed the amplitude $I$ of the current.  In the parametric
curves of $V_{y}(\phi )$ vs. $V_{x}(\phi )$ can we analyze the
breaking of rotational symmetry  in the different regimes of $I$ and $T$. 
In the case of rotational symmetry this kind of plot should
give a perfect circle. However, 
the square symmetry of the Josephson lattice 
will show up  in the shape of the curves.
In Ref.~\onlinecite{prbnos} the motion of single vortices
was studied. In this case,  the ``diagonal''
[11] direction is unstable against small changes in the angle
$\phi$, while the [10] and [01] directions are the 
preferred directions for vortex motion.\cite{prbnos}
This shows as ``horns'' in  parametric $V_y$ vs. $V_x$  plots,
which are finite segments of points lying in the $x$ or the 
$y$ axis. This implies the existance of transverse 
pinning in these directions 
(thus, it corresponds to orientational pinning). 
Here, we perform the same analysis for the different regimes
of the moving VL.

In Fig.19 we plot the voltages $V_{y}$ and  $V_{x}$ 
when varying the orientational angle
$\phi $ for different  current amplitudes $I$ 
and temperatures $T$.
In Fig.19(a) we have $I=0.06$ and $T=0.015$,  corresponding to the regime of  
a transversely pinned lattice. 
In this case most of the points are lying either on
the axis $V_{x}=0$ or on the axis $V_{y}=0$, 
indicating strong orientational pinning in the symmetry lattice
directions [10] and [01]. 
When increasing $T$ the orientational pinning
decreases and the length of the ``horns'' in the $x$ and $y$ axis decreases.
Fig.19(b) shows results for $I=0.06$ and $T=0.03$, which correspond
to $T>T_{tr}$ when there is a finite
transverse resistance. In this case the horns have disappeared
and orientational pinning is lost. 
However, the breaking of rotational symmetry
is still present in the star-shaped curves. 
Also in the high current regime, Fig.19(c), for a low temperature  $T=0.015$
($T<T_{tr}$) we see that there is orientational pinning with the presence of
horns, which again dissappear for $T>T_{tr}$ as it is shown in Fig.19(d) at
$T=0.035$. 
Finally for $T\gg T_M$, deep inside
in the liquid phase,  the stars tend to the circular shape 
of rotational invariance.

\section{DISCUSSION}

With all the information of the
previous sections we obtain the current-temperature phase diagram
which was advanced  in Fig.1.
For finite currents we have been able to identify
three different regimes,  a {\it pinned VL} for $T<T_p(I)$,  
a {\it transversely pinned VL} for $T_p(I)<
T < T_{tr}(I)$, and a {\it floating VL} for $T_{tr}(I) < T <
T_{M}(I)$. 
It is, however, difficult to define if the temperatures
$T_p, T_{tr}, T_M$ correspond to either
phase transitions or to dynamical crossovers.  
The comparisons we have made of the behavior of voltages (longitudinal and
transversal), structure factor and helicity modulus   show that
 something is happening at these temperatures.
Also, the
comparison for different system sizes of the behavior
of $S({\bf G})$ suggest  transitions for $T_p, T_{tr}, T_M$.
The transverse helicity modulus has strong fluctuations for 
$T_{tr}< T < T_{M}$. These  fluctuations are not reduced when
increasing the simulation time in a factor of $10$. This
could mean that actually the range of temperatures 
of $T_{tr}< T < T_{M}$ corresponds to a long crossover region towards
a liquid state.  Also, one may question if  the use of the helicity
modulus for these far from equilibrium states is correct, since $\Upsilon_x$
has been defined from the response of the equilibrium  free energy
to a twist in the boundary condition.
Most likely, the large fluctuations in $T_{tr}< T < T_{M}$ are due to
the very unstable and history-dependent steady states we find in this regime. 
For example the steady states obtained when increasing temperature from $T=0$
at fixed current differ from the steady states obtained when increasing
current at fixed temperature in this regime. They differ in the degree
of anisotropy and intensity of the Bragg peaks in the structure factor.
We think that this reflects the fact that the VL is unpinned in all
directions and the orientation of the moving state will depend on the
history for this regime. 

In any case, we can clearly  distinguish different dynamical  states
which define the regimes shown in the phase diagram of Fig.1. From the
experimental point of view these transitions (or crossovers) are
possible to measure. The depinning temperature $T_p$ can be obtained
from resistance measurements at finite currents. The transverse 
current-voltage characteristics and the transverse resistivity can be
measured for different longitudinal currents and temperatures, and therefore
 $T_{tr}$ could be obtained. The helicity modulus can be measured with the
two-coil technique. \cite{jjarev} It could be very interesting to see
the results of such measurements at finite currents. 
The structure factor and melting transition can not be measured
directly. However, in the presence of an external rf current, 
the dissappearance of Shapiro steps can indicate the melting
transition, \cite{harris} since they are
sensitive to the traslational order of  the vortex lattice.\cite{shapiro}
Therefore, we expect that the results obtained here could motivate
new experiments for fractional or submatching
fields in Josephson junction arrays as well as in other
superconductors with periodic pinning.

\acknowledgements
We acknowledge fruitful discussions with H. Pastoriza and J. V.
Jos\'{e}.
This work was supported by CONICET, CNEA, 
ANPCyT (PICT Nro. 03-00121-02151) and Fundaci\'{o}n Antorchas
(Proyecto Nro. 13532/1-96).

\appendix
\section{Periodic boundary conditions}
\subsection{Phases}

We want to obtain the periodic boundary condition (PBC) 
for superconducting phases $\theta(n_x,n_y)$.
In general we can write the PBC as:
\begin{eqnarray}
\theta({\bf n}+{\bf L_x})&=&\theta({\bf n})+u_x({\bf n})\nonumber\\
\theta({\bf n}+{\bf L_y})&=&\theta({\bf n})+u_y({\bf n})\;,
\end{eqnarray}
where ${\bf L_x}=(L_x,0)$ and ${\bf L_y}=(0,L_y)$. 
Taking into account that all variables should be independent of the
order of subsequent global translations and that the phases are defined
except for an addition of $2\pi l$, 
we are lead to the consistency condition
\begin{equation}
u_x({\bf n})+u_y({\bf n}+{\bf L_x})=
u_x({\bf n}+{\bf L_y})+u_y({\bf n})+2\pi l\;,
\end{equation}
Therefore, in order to specify the 
periodic boundary conditions we have to give the functions $u_x, u_y$, 
which will depend on the gauge for the vector potential ${\bf A}$.

The periodic boundary condition for the phases can be deduced
by requiring that all
physical quantities are invariant after a translation in the lattice
size. The supercurrents $S_\mu({\bf n})=
\sin[\Delta_\mu\theta({\bf n})-A_{\mu}({\bf n})]$ should satisfy:
\begin{eqnarray}
S_\mu({\bf n}+{\bf L_x})&=&S_\mu({\bf n})\nonumber\\
S_\mu({\bf n}+{\bf L_y})&=&S_\mu({\bf n})\;.
\end{eqnarray}
 This implies that the gauge invariant phase difference 
$\theta_\mu({\bf n})=\Delta_\mu\theta({\bf n})-A_{\mu}({\bf n})$ 
should satisfy
\begin{eqnarray}
\theta_\mu({\bf n}+{\bf L_x})&=&\theta_\mu({\bf n})+2\pi l\nonumber\\
\theta_\mu({\bf n}+{\bf L_y})&=&\theta_\mu({\bf n})+2\pi l'\;,
\end{eqnarray}
with $l, l'$ any integer. This condition leads to
\begin{eqnarray}
\Delta_\mu u_x({\bf n})&=&2\pi l 
+A_\mu({\bf n}+{\bf L_x})-A_\mu({\bf n})\nonumber\\
\Delta_\mu u_y({\bf n})&=&2\pi l' 
+A_\mu({\bf n}+{\bf L_y})-A_\mu({\bf n})\;.
\end{eqnarray}
We can choose the solution with $l=l'=0$.  
In the Landau gauge,  $A_x$ is a linear function of $n_y$ 
and $A_y$ is a linear function of $n_x$. 
Taking the origin such that $A_\mu({\bf n}=0)=0$, 
we obtain for the Landau gauge:
\begin{eqnarray}
 u_x({\bf n})&=&u_x(0)+A_y({\bf L_x})n_y\nonumber\\
 u_y({\bf n})&=&u_y(0)+A_x({\bf L_y})n_x\;.
\end{eqnarray}
The consistency condition (A2) requires
\begin{equation}
A_y({\bf L_x})L_y-A_x({\bf L_y})L_x=2\pi l\;.
\end{equation}
The term in the left side is equal to the total flux $2\pi f L_xL_y$,
therefore (A7) is equivalent to flux quantization, giving  
$f L_x L_y=N_v$, with $N_v=l$   the number of vortices.

If we take  $u_x(0)=u_y(0)=0$, we obtain for the PBC 
\begin{eqnarray}
\theta(n_x+L_x,n_y)&=&\theta(n_x,n_y)+A_y({\bf L_x})n_y  \nonumber \\
\theta(n_x,n_y+L_y)&=&\theta(n_x,n_y)+A_x({\bf L_y})n_x.
\end{eqnarray}
A particular choice can be the  gauge with 
$A_x({\bf n})=-2\pi f n_y$, $A_y({\bf n})=0$, which leads to Eq.~(7).

\subsection{External currents and electric fields}

In the presence of external currents or voltages the periodic boundary
conditions have to be reconsidered. 
In this case it is possible to have $\oint {\bf E}\cdot{\bf dl}\not =0$
in a path that encloses all the sample either in the $x$ or the $y$
direction. 
Therefore, in a closed path we have to consider the Faraday's law
$$ \oint {\bf E}\cdot{\bf dl} = -\frac{1}{c}\frac{d\Phi}{dt}.$$
The two-dimensional sample with PBC can be thought as
the surface of a torus in three-dimensions. The closed paths we
are considering are the two pathes that encircle the torus. 
The electric field is not a gradient of a potential, 
it is now given by ${\bf E}=-{\bf \nabla}V-\frac{1}{c}\frac{\partial {\bf
A}}{\partial t}$.  One possible solution is to consider a
vector potential $${\bf A}({\bf r},t)={\bf A_0}({\bf r})-{\bf\vec
\alpha}(t),$$ for which 
\begin{eqnarray}
{\bf H}_{\rm ext}&=&{\bf \nabla}\times{\bf A_0}\nonumber\\
{\bf E}_{\rm total}&=&\frac{1}{c}\frac{d{\bf\vec\alpha}}{dt}\nonumber
\end {eqnarray}

In our case, we take the adimensional vector potential as:
\begin{equation}
A_\mu({\bf n},t) = A_\mu^0({\bf n}) -\alpha_\mu(t),
\end{equation}
with $A_\mu^0({\bf n})$ in the Landau gauge  
($A_x^0({\bf n})=-2\pi f n_y$, $A_y^0({\bf n})=0$). Therefore 
the gauge invariant phase is:
$$\theta_\mu({\bf n},t)=\Delta_\mu\theta({\bf n},t)-A^0_\mu({\bf
n})+\alpha_\mu(t).$$
Then $\alpha_\mu$ acts as a global time-dependent phase in
the $\mu$ direction.

In the normalized units used in this paper, 
the electric field in the link defined by the junction ${\bf n}, \mu$ is 
$$E_\mu({\bf n})=-\Delta_\mu V({\bf n})- \frac{d A_\mu({\bf n})}{dt},$$
where the electrostatic potential is $V({\bf n})=-\frac{d\theta({\bf
n})}{dt}$. Therefore we have
$$E_\mu({\bf n})=\Delta_\mu\frac{d\theta({\bf
n})}{dt}+\frac{d\alpha_\mu}{dt}.$$
The average electric field in the $\mu$ direction is:
$$E^{\rm av}_\mu=\frac{1}{L_xL_y}\sum_{\bf n}E_\mu({\bf
n})=\frac{d\alpha_\mu}{dt},$$
where we have used the fact that $\sum_{\bf n}\Delta_\mu\frac{d\theta({\bf
n})}{dt}=0$
(which is the discrete equivalent of $\oint{\bf \nabla}V\cdot{\bf
dl}=0$).
The current in the link ${\bf n},\mu$ is, in normalized units,:
$$I_\mu({\bf n})=E_\mu({\bf n})+\tilde{S}_\mu({\bf n}),$$
with $\tilde{S}_\mu({\bf n})=S_\mu({\bf n})+\eta_\mu({\bf n},t)$.
Therefore the average current in the $\mu$ direction is
$$I^{\rm av}_\mu=\frac{1}{L_xL_y}\sum_{\bf n}I_\mu({\bf n})=E^{\rm
av}_\mu+S^{\rm av}_\mu+\eta^{\rm av}_\mu=
\frac{d\alpha_\mu}{dt}+\frac{1}{L_xL_y}\sum_{\bf n}\tilde{S}_\mu({\bf n})$$
There are two cases to consider: (i) {\it external current source}:
 the  external current is given
and the total voltage fluctuates and (ii) {\it external voltage source}:
the  external voltage is given and the total current fluctuates.

{\it(i) External current source:} the average current in
$\mu$ direction is fixed by the external current:
$I^{\rm av}_\mu=I_{\mu}^{\rm ext}$. 
The average electric field is a fluctuating quantity given by:
\begin{equation}
E^{\rm av}_\mu(t)=\frac{d\alpha_\mu}{dt}=I^{\rm ext}_\mu-\frac{1}{L_xL_y}
\sum_{\bf n}\tilde{S}_\mu({\bf n}).
\end{equation}
In this case, the $\alpha_\mu(t)$ is a dynamical variable, and its
time evolution is given by Eq.(A10).

{\it (ii) External voltage source:}  the average electric field
in the $\mu$ direction is fixed by the external electric field:
$E^{\rm av}_\mu=E_{\mu}^{\rm ext}$. 
Therefore, now the $\alpha_\mu$ is
given by the external source:
$$\alpha_\mu(t)= \int E_{\mu}^{\rm ext} dt,$$
and $\alpha_\mu(t)= E_{\mu}^{\rm ext} t$, 
if $E_{\mu}^{\rm ext}$ is time-independent. The average current is
now a fluctuating quantity given by:
$$I^{\rm av}_\mu=E_{\mu}^{\rm ext}+\frac{1}{L_xL_y}\sum_{\bf n}\tilde{S}_\mu
({\bf n}).$$

Let us see how the PBC are affected by a change
of gauge. The gauge transformations are the following:
\begin{eqnarray}
\theta({\bf n})\rightarrow \theta'({\bf n})&=&\theta({\bf
n})+\phi({\bf n})\nonumber\\
A_\mu({\bf n})\rightarrow A'_\mu({\bf n})&=&A_\mu({\bf
n})+\Delta_\mu\phi({\bf n})\nonumber\\
V({\bf n})\rightarrow V'({\bf n})&=&V({\bf
n})-\frac{d\phi({\bf n})}{dt}\nonumber
\end{eqnarray}
An interesting choice is:
$$\phi({\bf n})=\vec\alpha\cdot{\bf n}=\sum_\mu\alpha_\mu n_\mu$$
In this gauge we have $\theta_\mu({\bf n})=\Delta_\mu\theta({\bf
n})-A^0_\mu({\bf n})$ and the PBC for the phases is:
\begin{eqnarray}
\theta({\bf n}+{\bf L_x})&=&\theta({\bf n})+A_y({\bf
L_x})n_y+\alpha_xL_x  \nonumber \\
\theta({\bf n}+{\bf L_y})&=&\theta({\bf n})+A_x({\bf
L_y})n_x+\alpha_yL_y,
\end{eqnarray}
and for the voltages:
\begin{equation}
V({\bf n}+{\bf L_\mu})=V({\bf n})-\frac{d\alpha_\mu}{dt}L_\mu=
V({\bf n})-E^{\rm av}_\mu L_\mu
\end{equation}
The equations of motion in this gauge are:
\begin{eqnarray}
\frac{d\theta({\bf n})}{dt}&=&\sum_\mu\frac{d\alpha_\mu}{dt}
n_\mu-\sum_{\bf n'}G_{{\bf n},{\bf
n'}}\Delta_\mu\cdot\tilde{S}_\mu({\bf n'})\nonumber\\
\frac{d\alpha_\mu}{dt}&=&I^{\rm ext}_\mu-\frac{1}{L_xL_y}\sum_{\bf
n}\tilde{S}_\mu({\bf n})
\end{eqnarray}

The periodic boundary conditions with a fixed external current
using Eq.~(A10) was used previously in Ref.\onlinecite{minnh} for
$f=0$ (it was called a ``fluctuating twist boundary condition'') and
in Ref.\onlinecite{dd99} for $f\not=0$.  Also, the periodic
boundary conditions in the gauge of Eqs.~(A11-A12) were used
in Ref.~\onlinecite{vicente} for a time dependent $d$-wave
Ginzburg-Landau model.

\subsection{Helicity modulus}

The helicitity modulus $\Upsilon_\mu$ expresses the ``rigidity'' of the
system with respect to an applied ``twist'' in the periodic boundary
conditions. The twist $k_\mu$ is defined as a phase change of $L_\mu k_\mu$
between the two opposite boundaries which are connected through the PBC
in the $\mu$ direction 
$$\theta({\bf n}+{\bf L_\mu})=\theta({\bf
n})+L_\mu k_\mu.$$
The helicity modulus is obtained from the free energy $F(T,k_\mu)$   as:
\begin{equation}
\Upsilon_\mu=\left.\frac{1}{L^2}\frac{\partial^2F(T,k_\mu)}{\partial
k_\mu^2}\right|_{k_\mu=0}
\end{equation}
It is clear from Eq.~(A11) that $k_\mu=\alpha_\mu$. Then, in order
to evaluate the helicity modulus, $\alpha_\mu$ must be set to zero.
This means that the helicity modulus can not be calculated in
the direction in which there is an applied current, since
it gives a fluctuating twist $k_\mu(t)$.

\subsection{Algorithm}

The core of the numerical calculation is to invert the Eq.~(9). 
This means to solve a discrete Poisson equation of the form,
\begin{equation}
\Delta_\mu^2f({\bf n})=d({\bf n})
\label{poiss}
\end{equation}
with the periodic boundary conditions
\begin{eqnarray}
f({\bf n}+{\bf L_x})&=&f({\bf n})\nonumber\\
f({\bf n}+{\bf L_y})&=&f({\bf n})\;.
\end{eqnarray}
The linear system of $L_xL_y$ equations of (\ref{poiss}) is singular.
Physically, the reason is that in Eq.(\ref{dyn0}),  $f({\bf
n})=d\theta({\bf n})/dt$ 
corresponds to a voltage, which is defined 
except for a constant.  We choose the voltage reference such
that it has zero mean:
\begin{equation}
\sum_{\bf n} f({\bf n})=0\;,
\end{equation}
other choices are also possible (like, for example, fixing $f({\bf
n_0})=0$ at  a given site ${\bf n}_0$).

The method we use to invert the Eq.~(\ref{poiss})  is based on 
the Fourier Accelerated and Cyclic Reduction (FACR) 
algorithm. \cite{recipes}
In this case, we take first the discrete Fourier transform
in the $x$ direction:
\begin{equation}
\tilde{d}(k_x,n_y)=\sum_{n_x} d(n_x,n_y)e^{i\frac{2\pi k_x n_x}{L_x}}\;,
\end{equation}
which with a fast Fourier algorithm takes a computation time of 
order $L_x\log L_x$.
This leads to the following equation:
\begin{equation}
\epsilon_{k_x}\tilde{f}(k_x,n_y)-
\tilde{f}(k_x,n_y-1)-\tilde{f}(k_x,n_y+1)=\tilde{d}(k_x,n_y)\;,
\end{equation}
with $\epsilon_{k_x}=4-2\cos\frac{2\pi k_x}{L_x}$
and with boundary condition $\tilde{f}(k_x,L_y+1)=\tilde{f}(k_x,1)$.
This is a cyclic tridiagonal equation which can be solved with
a simple LU decomposition algorithm in a computation time
of order $L_y$.\cite{recipes} In this way the 
$\tilde{f}(k_x,n_y)$ is obtained from (A19). Finally we take
the inverse Fourier transform to obtain
\begin{equation}
f(n_x,n_y)=\frac{1}{L_x}\sum_{k_x} \tilde{f}(k_x,n_y)e^{-i\frac{2\pi k_x
n_x}{L_x}}\;.
\end{equation}
This algorithm takes a computation time which is of order
$L_xL_y(\log L_x + A)$ with the constant $A \sim 1 $. This is faster
than the two dimensional Fourier transform method used
by Eikmans and van Himbergen,\cite{eik} which takes a computation
time of order $L_xL_y(\log L_x + \log L_y)$.

\begin{figure}[tbp]
\centerline{\epsfxsize=8.5cm \epsfbox{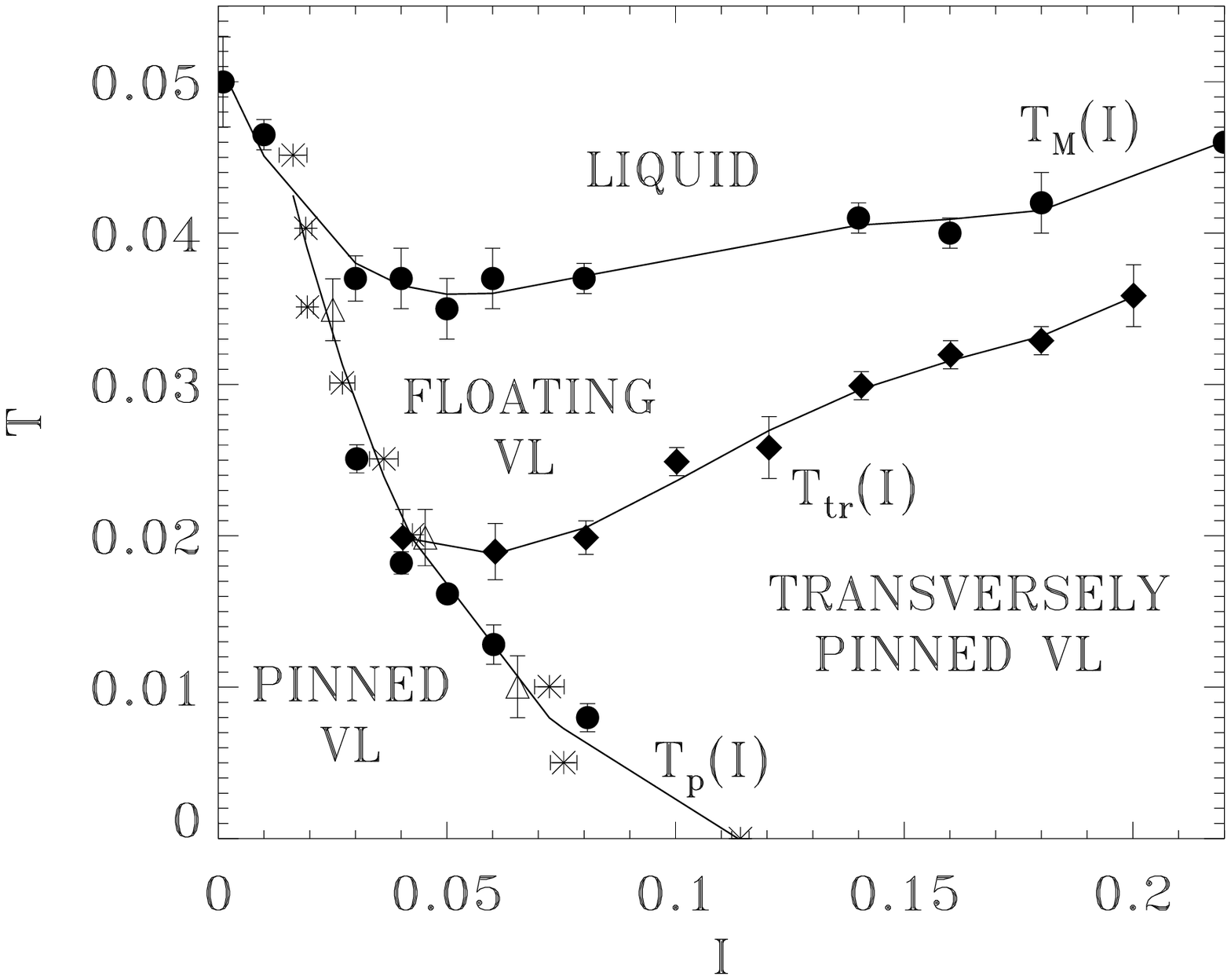}}
\caption{ $I-T$ Phase diagram for $f=1/25$. $T_M(I)$ line is obtained from
$S({\bf G})$ vs. $T$ curves $(\bullet)$. $T_p(I)$ line is obtained from IV
curves $(\star)$, from $S({\bf G})$ vs. $T$ curves $(\bullet)$ and from
$\langle V_y\rangle$
vs. $T$ curves $(\triangle)$. $T_{tr}(I)$ curve is obtained from
$\langle V_{tr}\rangle$ vs.
$T$ curves $(\blacklozenge)$. 
}
\label{fig1}
\end{figure}

\begin{figure}[tbp]
\centerline{\epsfxsize=8.5cm \epsfbox{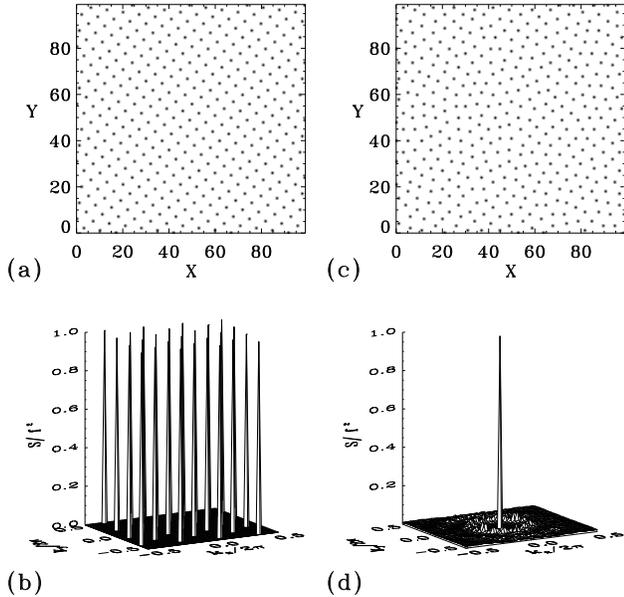}}
\caption{Vortex configuration for a low current $I=0.01 \ll I_c(0)$ 
and a low temperature, $T=0.01 < T_M$: 
(a) tilted square vortex lattice, oriented in the $[4a,3a]$ direction
and (b) corresponding structure factor $S({\bf k})$.
At high temperature, $T=0.05 \gtrsim T_M$:
(c) disordered vortex array,
(d) vortex liquid-like structure factor.
}
\label{fig2}
\end{figure}

\begin{figure}[tbp]
\centerline{\epsfxsize=8.5cm \epsfbox{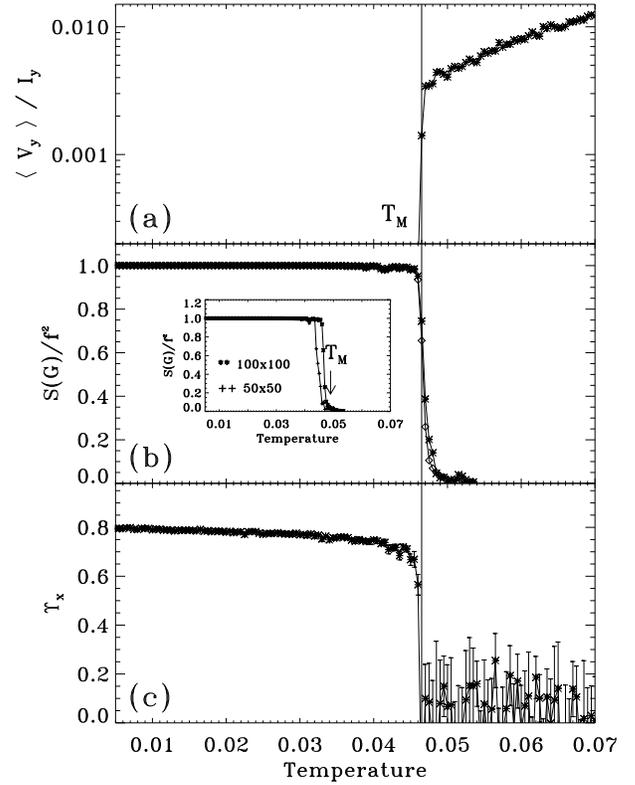}}
\caption{For low current $I \ll I_c(0)$, $I=0.01$:
(a) $<V_y>/I_y$ vs. $T$,
(b) $S({\bf G_1})$ $(\diamond)$ and $S({\bf G_2})$ $(\star)$ vs. $T$,
inset: size effect in $S({\bf G_1})$,
(c) $\Upsilon_x$ vs. $T$.
}
\label{fig3}
\end{figure}

\begin{figure}[tbp]
\centerline{\epsfxsize=8.5cm \epsfbox{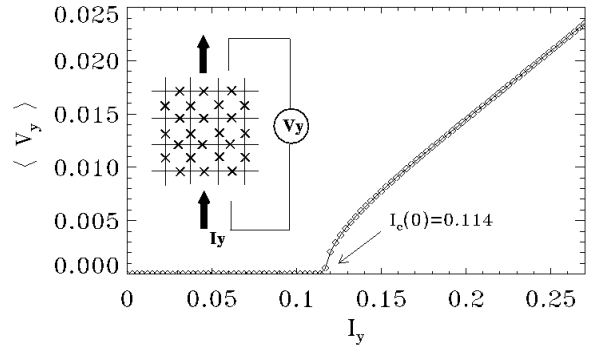}}
\caption{Zero temperature  $IV$ curve. There is a 
critical current $I_c(0)=0.114\pm0.002$.}
\label{fig4}
\end{figure}

\begin{figure}[tbp]
\centerline{\epsfxsize=8.5cm \epsfbox{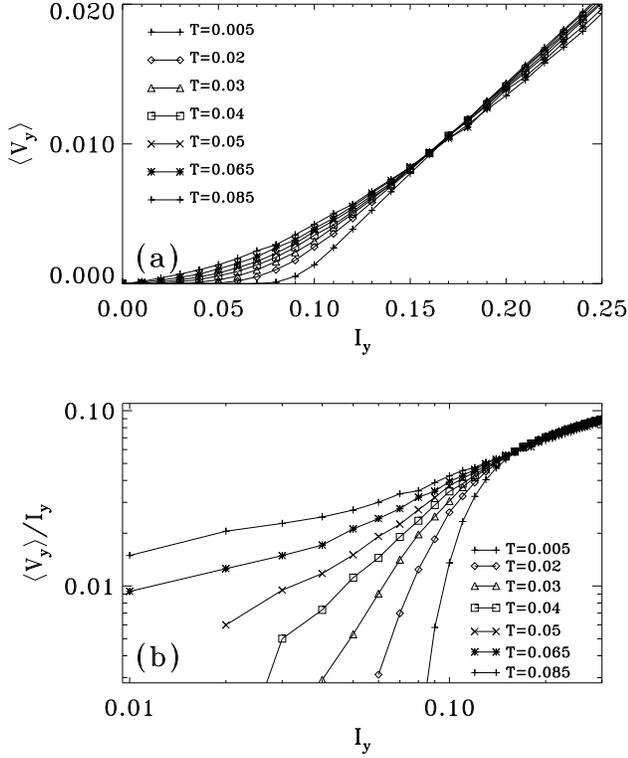}}
\caption{$IV$ curves for different temperatures.
(a) Linear scale, crossing point at $I^*=0.165$.
(b) Log-log plot of $R=V/I$ vs. $I$ for the same temperatures.  
}
\label{fig5}
\end{figure}

\begin{figure}[tbp]
\centerline{\epsfxsize=8.5cm \epsfbox{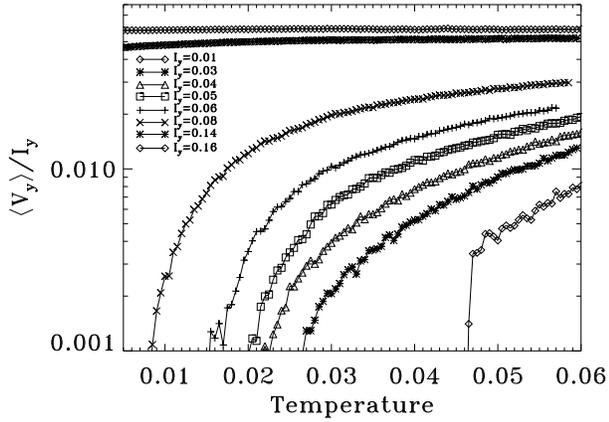}}
\caption{$<V_y>/I_y $ vs temperature curves for different dc currents ($I_y$).  
}
\label{fig6}
\end{figure}

\begin{figure}[tbp]
\centerline{\epsfxsize=8.5cm \epsfbox{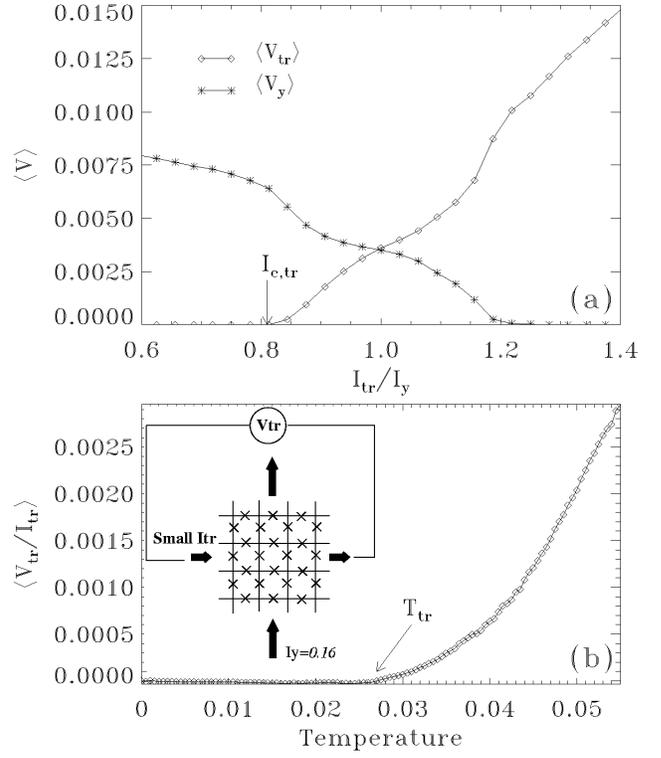}}
\caption{Transverse depinnig temperature determinations: (a) $Tr-IV$ curve
and $<V_y>$ vs $I_{tr}/I_y$  at $T=0$ and high longitudinal current,
$I_y=0.16$. (b) With small transverse current applied, $I_x=0.01$, and
$I_y=0.16$,  $<V_{tr}/I_{tr}>$ vs $T$.       
}
\label{fig7}
\end{figure}

\begin{figure}[tbp] 
\centerline{\epsfxsize=8.5cm \epsfbox{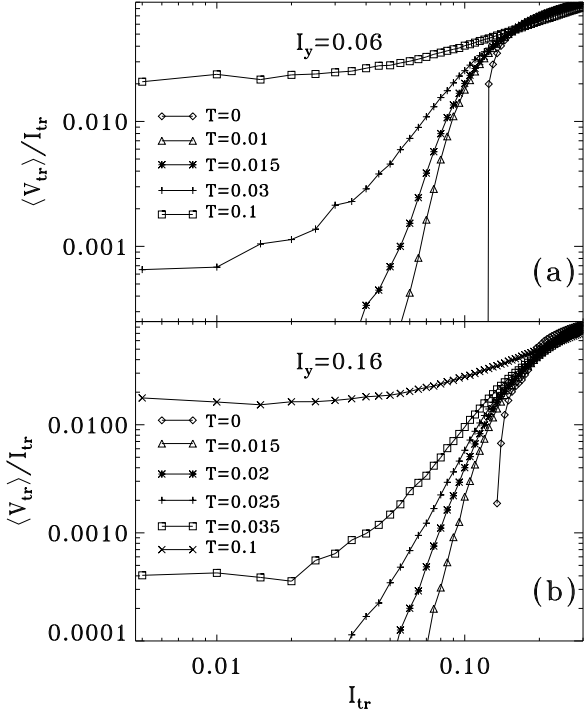}}
\caption{Transverse $IV$ curves for different temperatures:
(a) low current  $I_y < I_c(0)$, $I_y=0.06$, 
(b) high  current  $I_y > I_c(0)$, $I_y=0.16$.        
}
\label{fig8}
\end{figure}

\begin{figure}[tbp]
\centerline{\epsfxsize=8.5cm \epsfbox{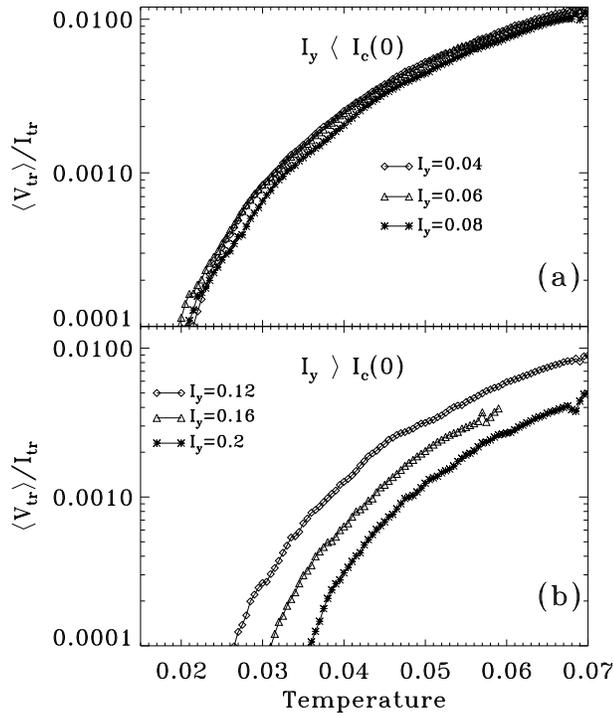}}
\caption{$V_{tr}/I_{tr}$ vs $T$ curves for different dc currents applied in the 
$y$ direction and small transverse current applied, $I_x=0.01$:
(a) low current  $I_y < I_c(0)$,
(b) high  current $I_y > I_c(0)$.        
}
\label{fig9}
\end{figure}

\begin{figure}[tbp]
\centerline{\epsfxsize=8.5cm \epsfbox{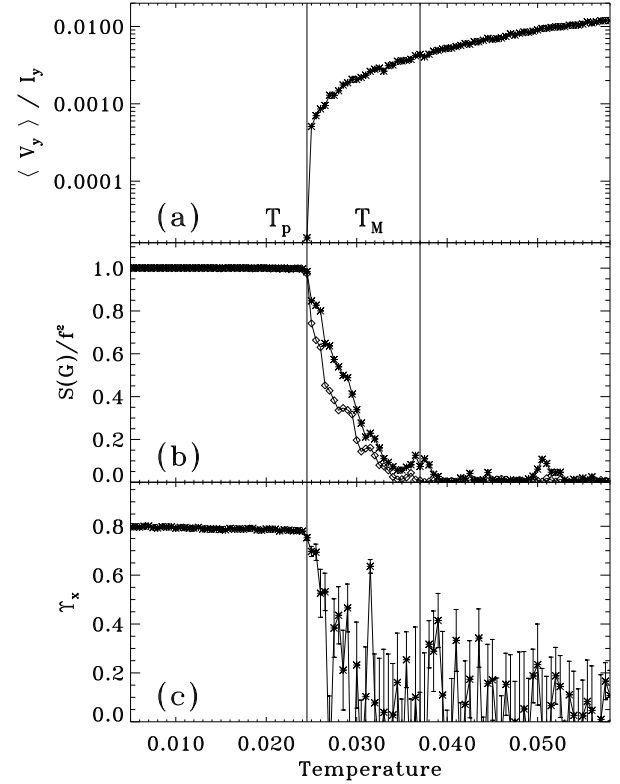}}
\caption{For low current $I< I_c(0)$, $I=0.03$:
(a) $<V_y>/I_y$ vs. $T$,
(b) $S({\bf G_1})$ $(\diamond)$ and $S({\bf G_2})$ $(\star)$ vs. $T$,
(c) $\Upsilon_x$ vs. $T$.
}
\label{fig10}
\end{figure}

\begin{figure}[tbp]
\centerline{\epsfxsize=8.5cm \epsfbox{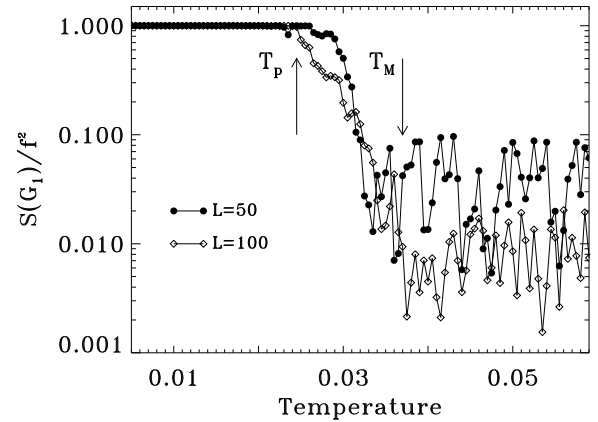}}
\caption{For low current $I< I_c(0)$, $I=0.03$: size effect in $S({\bf G_1})$.
}
\label{fig11}
\end{figure}

\begin{figure}[tbp]
\centerline{\epsfxsize=8.5cm \epsfbox{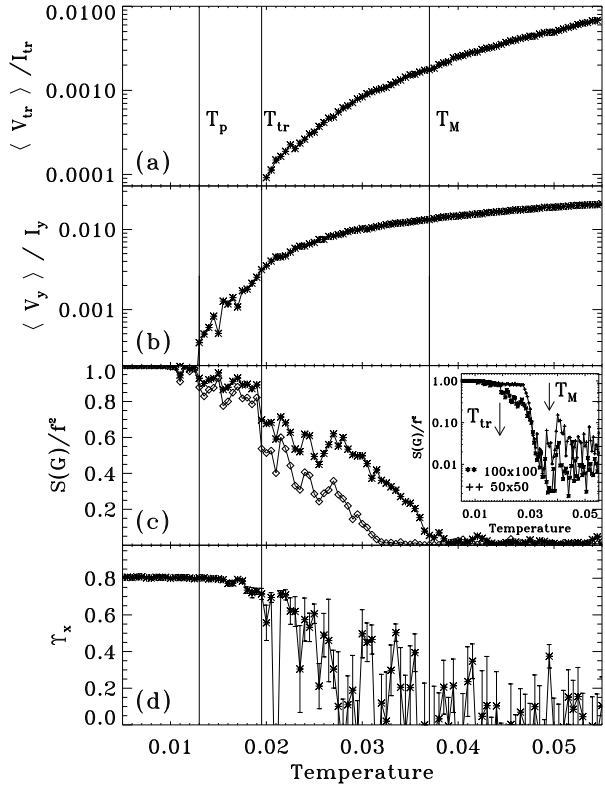}}
\caption{For $I< I_c(0)$, $I=0.06$:
(a) $<V_{tr}>/I_{tr}$ vs. $T$,
(b) $<V_y>/I_y$ vs. $T$,
(c) $S({\bf G_1})$ $(\diamond)$ and $S({\bf G_2})$ $(\star)$ vs. $T$,
inset: size effect in $S({\bf G_1})$,
(d) $\Upsilon_x$ vs. $T$.
}
\label{fig12}
\end{figure}

\begin{figure}[tbp]
\centerline{\epsfxsize=8.5cm \epsfbox{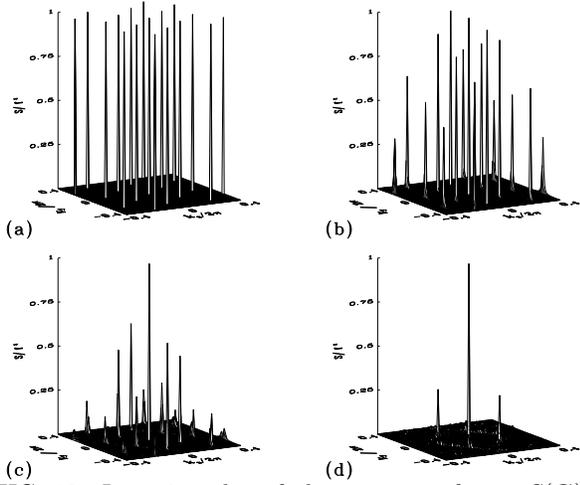}}
\caption{Intensity plot of the structure factor $S({\bf G})$ at $I<I_c(0)$,
 $I=0.06$ for different temperatures:
(a) $T < T_p$, $T=0.0005$,
(b) $T_p < T < T_{tr}$, $T=0.015$,
(c) $T_{tr} < T < T_M$, $T=0.025$,
(d) $T \leq T_M$, $T=0.034$.         
}
\label{fig13}
\end{figure}

\begin{figure}[tbp]
\centerline{\epsfxsize=8.5cm \epsfbox{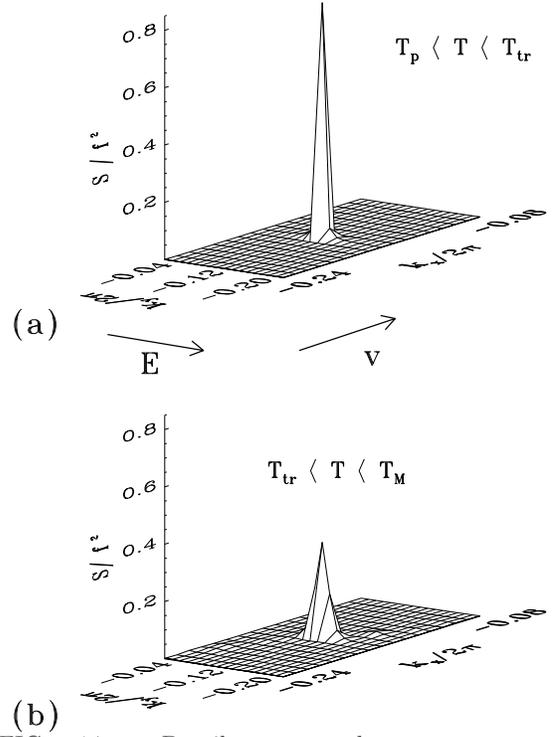}}
\caption{Details on the structure factor peak, 
${\bf G_1}=\frac{2\pi}{a}(-4/25,-3/25)$, at $I<I_c(0)$,
 $I=0.06$, for different temperatures:
(a) $T_p < T < T_{tr}$, $T=0.018$,
(b) $T_{tr} < T < T_M$, $T=0.023$.        
}
\label{fig14}
\end{figure}

\begin{figure}[tbp]
\centerline{\epsfxsize=8.5cm \epsfbox{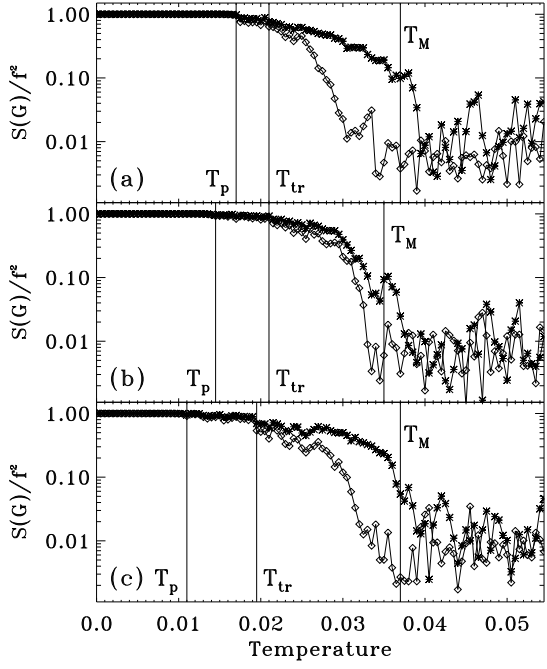}}
\caption{$S({\bf G_1})$ $(\diamond)$ and $S({\bf G_2})$ $(\star)$ vs. $T$
 for different dc current applied in $y$ direction:
(a) $I_y=0.04$,
(b) $I_y=0.05$,
(c) $I_y=0.06$.        
}
\label{fig15}
\end{figure}

\begin{figure}[tbp]
\centerline{\epsfxsize=8.5cm \epsfbox{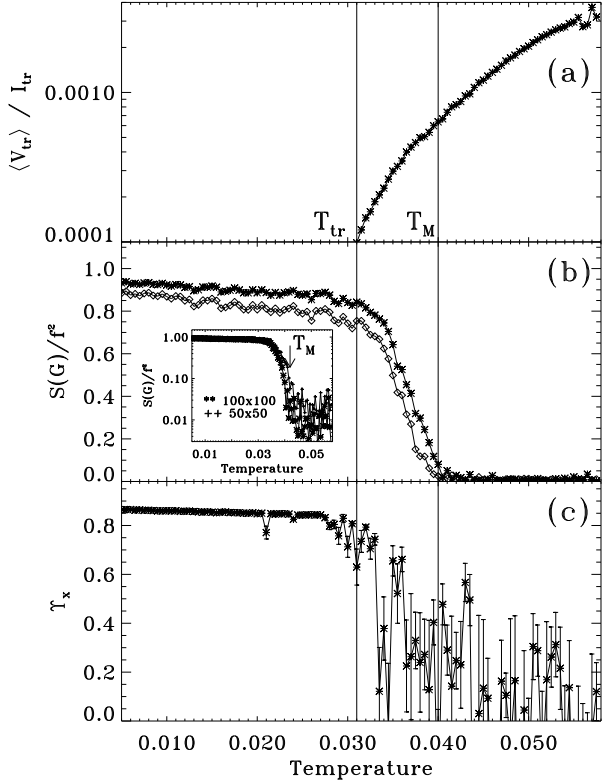}}
\caption{For high current $I> I_c(0)$, $I=0.16$:
(a) $<V_{tr}>/I_{tr}$ vs. $T$,
(b) $S({\bf G_1})$ $(\diamond)$ and $S({\bf G_2})$ $(\star)$ vs. $T$,
inset: size effect in $S({\bf G_2})$,
(c) $\Upsilon_x$ vs. $T$.
}
\label{fig16}
\end{figure}

\begin{figure}[tbp]
\centerline{\epsfxsize=8.5cm \epsfbox{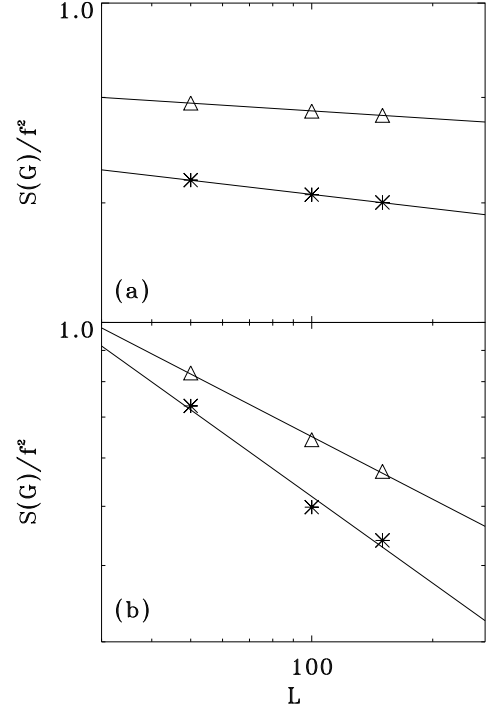}}
\caption{Finite size
analysis and power law fit of $S({\bf G})\sim L^{-\eta_G}(I,T)$ for a moving
vortex lattice at $I>I_c(0)$, $I=0.16.$ We obtain: (a) for  $T < T_{tr}$,
$T=0.02$,  $\eta_{G_1}=0.023$ $(\star)$ and $\eta_{G_2}=0.013$, $(\triangle)$,
(b) for $T_{tr} < T < T_M$, $T=0.035$, $\eta_{G_1}=0.471$ $(\star)$ and 
$\eta_{G_2}=0.34$ $(\triangle).$   
}
\label{fig17}
\end{figure}

\begin{figure}[tbp]
\centerline{\epsfxsize=8.5cm \epsfbox{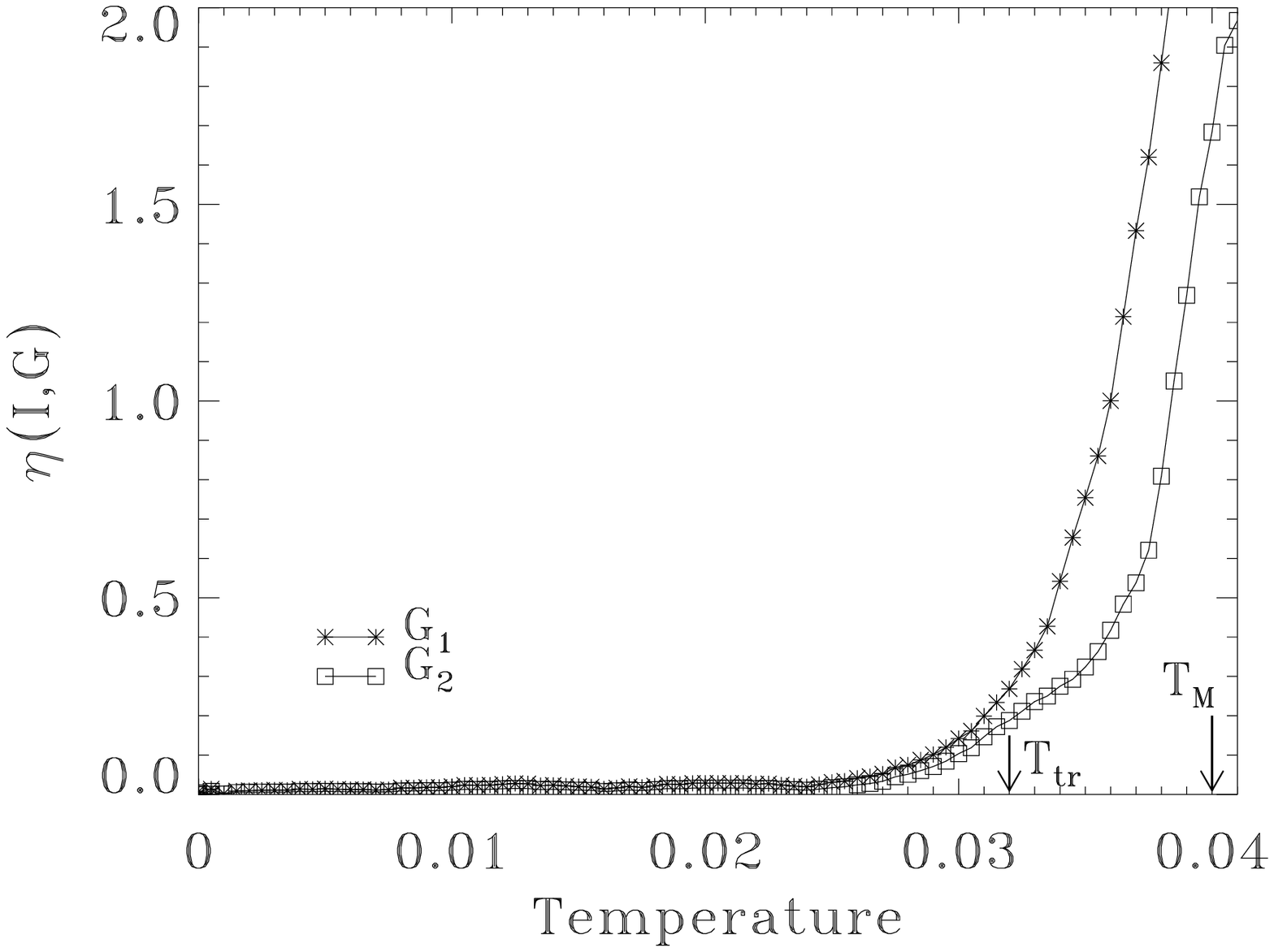}}
\caption{At $I=0.16$, $\eta(I,{\bf G})$ vs $T$, for ${\bf G_1}$
 and ${\bf G_2}$.     
}
\label{fig18}
\end{figure}

\begin{figure}[tbp]
\centerline{\epsfxsize=8.5cm \epsfbox{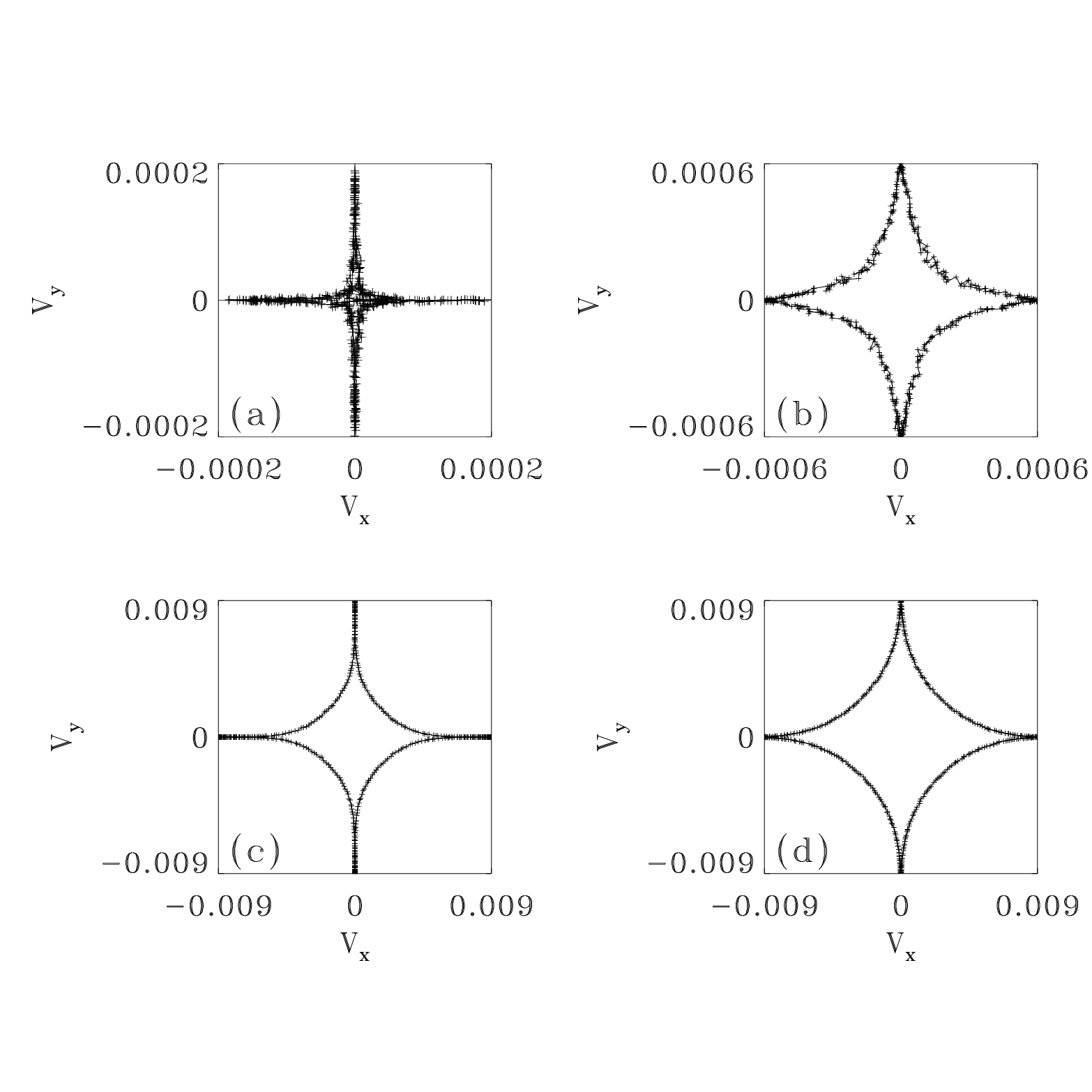}}
\caption{Parametric curves $V_y(\phi)$ vs. $V_x(\phi)$ for different dc current
applied and temperatures:
(a) $|I_y|=0.06$ and $T=0.015$,
(b) $|I_y|=0.06$ and $T=0.03$,
(c) $|I_y|=0.16$ and $T=0.015$,
(d) $|I_y|=0.16$ and $T=0.035$.        
}
\label{fig19}
\end{figure}


\begin{references}



\bibitem{martinoli}  P. Martinoli, O. Daldini, C. Leemann, and E. Stocker,
Solid State Comm. {\bf 17}, 205 (1975); P. Martinoli, O. Daldini, C.
Leemann, and B. Van den Brandt, Phys. Rev. Lett. {\bf 36}, 382 (1976). 


\bibitem{wire}  B. Pannetier, J. Chaussy, R. Rammal, and J. C. Villegier,
Phys. Rev. Lett. {\bf 53}, 1845 (1984); H. D. Hallen {\it et al.},
Phys. Rev. Lett. {\bf 71}, 3007 (1993); K. Runge and B. Pannetier,
Europhys. Lett. {\bf 24}, 737 (1993).



\bibitem{jjarev} For recent reviews in Josephson arrays see
R. S. Newrock, C. J. Lobb, U. Geigenmuller and M. Octavio, 
Solid State Physics {\bf 54}, 263 (2000); 
{\it Lectures on superconducting networks and mesoscopic systems},
Ed. C. Giovanella and C. J. Lambert (American Institute of Physics,
AIP Conf. Proc. 427, New York, 1998)
See also J. C. Ciria and C. Giovanella, J. Phys. Condes. Matter {\bf 10}, 1453
(1998);
{\it Macroscopic Quantum Phenomena and Coherence in Superconducting Networks},
Ed. C. Giovanella and M. Tinkham (World Scientific, Singapore, 1995) and
{\it Proceedings of the ICTP Workshop on Josephson-junction arrays}, Physica
B {\bf 222}, 253 (1996).

\bibitem{jjpin}  C. J. Lobb, D. W. Abraham. and M. Tinkham, Phys. Rev. B 
{\bf 27}, 150 (1983).


\bibitem{jjafrus}  See for example M. S. Rzchowski, S. P. Benz, M. Tinkham, 
and C. J. Lobb, Phys. Rev. B {\bf 42}, 2041 (1990). 

\bibitem{dot}  J. I. Mart\'{\i}n, M. V\'elez, J. Nogu\'es, and Ivan K.
Schuller , Phys. Rev. Lett. {\bf 79}, 1929 (1997);
D. J. Morgan and J. B. Ketterson, Phys. Rev. Lett. {\bf 80}, 3614
(1998); Y. Jaccard {\it et al.}, Phys. Rev. B {\bf 58}, 8232 ( 1998);
M.J. Van Bael {\it et al.}, Phys. Rev. B {\bf 59}, 14674 (1999);
J.I.Mart\'{\i}n {\it et al.}, Phys. Rev. Lett. {\bf 83}, 1022 (1999);
T. Puig {\it et al.}, Phys. Rev. B {\bf 58}, 5744 (1998).

\bibitem{hole}
M. Baert, V. V. Metlushko, R. Jonckheere, V. V. Moshchalkov,
and Y. Bruynseraede, Phys. Rev. Lett {\bf 74}, 3269 (1995);
K. Harada, O. Kamimura, H. Kasai, T. Matsuda, A. Tonomura
and  V. V. Moshchalkov, Science {\bf 271}, 1393 (1996);
J.Y. Lin {\it et al.},  Phys. Rev. B {\bf 54}, R12714 (1996);
A. Bezryadin, Yu. B. Ovchinnikov, and B. Pannetier,
Phys. Rev. B {\bf 53}, 8553 (1996);
E. Rossel {\it et al.}, Phys. Rev. B {\bf 53}, R2983;
A. Castellanos, R. Wondenweber, G. Ockenfuss, A. v.d. Hart, 
and K. Keck, Appl. Phys. Lett. {\bf 71}, 962 (1997)
V. V. Moshchalkov, M. Baert, V. V. Metlushko, E. Rosseel, M.
J. van Bael, K. Temst, Y Bruynseraede, and R. Jonckheere, 
Phys. Rev. B {\bf 57}, 3615 (1998);
V. Metlushko {\it et al.}, Phys. Rev. B {\bf 59}, 603 (1999);
V. Metlushko {\it et al.}, Phys. Rev. B {\bf 60}, 12585 (1999);
L. V. Look {\it et al.}, Phys. Rev. B {\bf 60}, R6998 (1999). 


\bibitem{sfield}  S. B. Field, S. S. James, J. Barentine, V. Metlushko, G.
Crabtree, H. Shtrikman, B. Ilic, and S. R. J. Brueck,  preprint
cond-mat/003415.

\bibitem{yan} Y. Fasano, J. A. Herbsommer, F. de la Cruz, F. Pardo, 
P. L. Gammel, E. Bucher and D. J. Bishop, Phys. Rev. B {\bf 60}, R15047 (1999).

\bibitem{martinoli0} P. Martinoli, Phys. Rev. B {\bf 17}, 1175 (1978).

\bibitem{PT} V. L. Pokrovskii and A. L. Talanov, Sov. Phys. JETTP
{\bf 51}, 134 (1980).

\bibitem{martinoli2} P. Martinoli, M. Nsabimana, G. A. Racine and H.
Beck, Hel. Phys. Acta {\bf 55}, 655 (1982).

\bibitem{nelson} 
D. R. Nelson in {\it Phase Transitions and
Critical Phenomena}, edited by C. Domb and J. L. Lebowitz
(Academic, New York, 1983), Vol. 7.

\bibitem{franz}  M. Franz and S. Teitel, Phys. Rev. Lett. {\bf 73}, 480
(1994) and Phys. Rev. B {\bf 51}, 6551 (1995).

\bibitem{hattel} S. Hattel and J. M. Wheatley,Phys. Rev. B {\bf 51},
 11951 (1995).

\bibitem{nori0}  C. Reichhardt, C. J. Olson and F. Nori, Phys. Rev. B {\bf 54%
}, 16108 (1996); Phys. Rev. B {\bf 57}, 7937 (1998).

\bibitem{carneiro0} W. A. M. Morgado and G. Carneiro, Physica C {\bf
238}, 195 (1998).

\bibitem{nori}  C. Reichhardt, C. J. Olson and F. Nori, Phys. Rev. Lett. 
{\bf 78}, 2648 (1997); Phys. Rev. B {\bf 58}, 6534 (1998).

\bibitem{nori2}  C. Reichhardt and F. Nori, Phys. Rev. Lett. {\bf 82}, 414
(1999).

\bibitem{md99}  V. I. Marconi and D. Dom\'{\i }nguez, Phys. Rev. Lett {\bf %
82, }4922 (1999).

\bibitem{carneiro99} G. Carneiro, J. Low Temp. Phys. {\bf 177}, 1323
(1999).

\bibitem{reich}  C. Reichhardt, G. Zimanyi, Phys. Rev. B
{\bf 61}, 14354 (2000).

\bibitem{carneiro} G. Carneiro, preprint.


\bibitem{prbnos} V. I. Marconi, S. Candia, P. Balenzuela, H. Pastoriza, D.
Dom\'{\i}nguez and P. Martinoli,  Phys. Rev. B (01 August 2000).

\bibitem{KV} A. E. Koshelev and V. M. Vinokur, Phys. Rev. Lett.
{\bf 73}, 3580  (1994);
S. Scheidl and V. M. Vinokur, Phys. Rev. B {\bf 57}, 13800
(1998).

\bibitem{gld}  T. Giamarchi and P. Le Doussal, Phys. Rev. Lett. {\bf 76},
3408 (1996); P. Le Doussal and T. Giamarchi, Phys. Rev. B {\bf 57}, 11356
(1998).

\bibitem{bmr}L. Balents, M. C. Marchetti and L. Radzihovsky, Phys. Rev. B 
{\bf 57}, 7705 (1998).

\bibitem{pardo} F. Pardo {\it et al},  Phys. Rev. Lett. {\bf 78}, 4633 (1997);
and Nature {\bf 396}, 398 (1998).

\bibitem{simu} K. Moon {\it et al.}, Phys. Rev. Lett. {\bf 77}, 2778
(1996); S. Ryu {\it et al.}, Phys. Rev. Lett. {\bf 77}, 5114 (1996);
S. Spencer and H. J. Jensen, Phys. Rev. B {\bf 55}, 8473
(1997); C. J. Olson {\it et al.}, Phys. Rev. Lett. {\bf 81}, 3757
(1998); A. B. Kolton, D. Dom\'{\i}nguez, N. Gronbech-Jensen,
Phys. Rev. Lett. {\bf 83}, 3061 (1999); A. B. Kolton {\it et al.},
cond-mat/0004059.

\bibitem{dgb} D. Dom\'{\i}nguez, N. Gr\o nbech-Jensen and A.R. Bishop,
Phys. Rev. Lett. {\bf 78}, 2644 (1997).

\bibitem{dd99} D. Dom\'{\i }nguez, Phys. Rev. Lett. {\bf 82}, 181 (1999).

\bibitem{dyna0}  J.\ S.\ Chung, K.\ H.\ Lee, and D.\ Stroud, Phys.\ Rev.\ B 
{\bf 40}, 6570 (1989), K. K. Mon and S. Teitel, Phys. Rev. Lett. {\bf
63}, 673 (1989).
 
\bibitem{falo} F.\ Falo, A.\ R.\ Bishop, and S.\ P.\ Lomdahl, Phys.\ Rev.\ B
 {\bf 41}, 10983(1990).

\bibitem{eik} H. Eikmans and J.E. van Himbergen, Phys. Rev. B {\bf 41},
8927 (1990).

\bibitem{acvs}  D. Dom\'{\i }nguez, J. V. Jos\'{e}, A. Karma and C. Wiecko, 
Phys. Rev. Lett. (1991) {\bf 67}, 2367; 
D. Dom\'{\i }nguez, Phys. Rev. Lett. {\bf 72}, 3096 (1994).

\bibitem{kim93} S. Kim and M. Y. Choi, Phys. Rev. B {\bf 48}, 322
(1993).


\bibitem{vorstroud} W. Yu, K. H. Lee, and D. Stroud, Phys. Rev. B
{\bf 47}, 5906 (1993).

\bibitem{vorjose}  T. J. Hagenaars, P.H.E. Tiesinga, 
J.E. van Himbergen and
J.V. Jos\'{e}, Phys. Rev. B {\bf 50}, 1143 (1994).



\bibitem{dj96} D. Dom\'{\i }nguez and J. V. Jos\'{e},
Phys. Rev. B. {\bf 53}, 11692 (1996).



\bibitem{teitel25} Y. H. Li and S. Teitel, 
Phys. Rev. B {\bf 47}, 359 (1993).

\bibitem{hellerq} M. C. Hellerqvist {\it et al.}, Phys. Rev. Lett. 
{\bf 76}, 4022  (1996).

\bibitem{stroud99}  K. D. Fisher, D. Stroud and L. Janin, preprint
cond-mat/9906068.

\bibitem{choi} M. Yoon, M. Y. Choi and B. J. Kim, Phys. Rev. B
{\bf 61}, 3263 (2000).



\bibitem{vicente}  J. J. Vicente Alvarez, D. Dom\'{\i }nguez and C. A.
Balseiro, Phys. Rev. Lett. {\bf 79}, 1373 (1997).

\bibitem{minnh} B. J. Kim, P. Minnhagen, and P. Olsson, Phys. Rev. B
{\bf 59}, 11506 (1999).

\bibitem{recipes} W. H. Press, S. A. Teukolsky, W. T. Vetterling
and B. P. Flannery, {\it Numerical Recipes} (Cambridge University Press,
New York, 1992). 


\bibitem{harris} Harris {\it et al.}, Phys. Rev. Lett. {\bf } (1995).

\bibitem{shapiro} V. I. Marconi and D. Dom\'{\i}nguez, unpublished.

\end{references}
\end{document}